\documentclass[journal=jacsat,manuscript=article]{achemso}
\documentclass[main.tex]{subfiles}

\begin{document}
%%%%%%%%%%%%%%%%%%%%%%%%%%%%%%%%%%%%%%%%%%%%%%%%%%%%%%%%%%%%%%%%%%%%%
%% The manuscript does not need to include \maketitle, which is
%% executed automatically.  The document should begin with an
%% abstract, if appropriate.  If one is given and should not be, the
%% contents will be gobbled.
%%%%%%%%%%%%%%%%%%%%%%%%%%%%%%%%%%%%%%%%%%%%%%%%%%%%%%%%%%%%%%%%%%%%%
\begin{abstract}
 
The microscopic description of the local structure of water remains an open challenge. Here, we adopt an agnostic approach to understanding water's hydrogen bond network using data harvested from molecular dynamics simulations of an empirical water model. A battery of state-of-the-art unsupervised data-science techniques are used to characterize the free energy landscape of water starting from encoding the water environment using local-atomic descriptors, through dimensionality reduction and finally the use of advanced clustering techniques. Analysis of the free energy at ambient conditions was found to be consistent with a rough single basin and independent of the choice of the water model. We find that the fluctuations of the water network occur in a high-dimensional space which we characterize using a combination of both atomic descriptors and chemical-intuition based coordinates. We demonstrate that a combination of both types of variables are needed in order to adequately capture the complexity of the fluctuations in the hydrogen bond network at different length-scales both at room temperature and also close to the critical point of water. Our results provide a general framework for examining fluctuations in water under different conditions.

 \end{abstract}

%\alex{TODO list: 
%\begin{itemize}
%    \item Check empty cites and look in bibliography to fill them.
  %  \item Adapt last paragraph of the introduction to the real structure of the paper.
   % \item \deleted{Add SOAP description  based on agreed nomenclature}
 %   \item Fill up the 's at various points in the manuscript.
 %   \item Check captions.
 %   \item Check repeated labels.
 %   \item Fix SI to be a real document and Fix references in Main Doc to the SI figures. 
%\end{itemize}
%}

\section{Introduction}

Water is one of the most ubiquitous solvents found in numerous physical, chemical, biological and even technological contexts\cite{franks2000water,ball2008water,evans1996water}. Unlike most simple liquids, it is characterized by various thermodynamic and dynamical anomalies, such as an enhancement in its compressibility upon supercooling.
\cite{chaplin2019structure,debenedetti2003supercooled,angell2004amorphous}. 
%\cite{chaplin2019structure,debenedetti2003supercooled,angell2004amorphous,Introduction to phase transitions and critical phenomena}. 
Its unique properties are rooted in its ability to form a directed hydrogen bond network\cite{smith2004energetics,geiger1982low}. Despite long study, the origins of these anomalies in terms of the underlying structure and dynamics of hydrogen bonds continues to be a topic of lively discussion\cite{characterizingdeboue,chaplin2019structure,yeh1999orientational,rao2010structural}. 

Perhaps one of the most debated issues in aqueous sciences over the last two decades, has been how its instantaneous local structure evolves as a function of temperature and pressure\cite{holten2014two}. Over a century ago, Roentgen proposed that the anomalies of water could be rationalized as a competition between two different water structures namely, a low-density-liquid (LDL) and a high-density-liquid (HDL) \cite{rontgen1892ueber}. Since then, various experiments\cite{nilsson2012fluctuations,taschin2013evidence,wernet2004structure} as well as numerical simulations \cite{soper2019water,pettersson2018two,geiger1982low,gallo2016water} have attempted to provide a microscopic picture of the structure of water leading to contradictory conclusions. Very recently, there have been several reports proving the existence of a second critical point in water under supercooling with classical atomistic models\cite{debenedetti2020second,gartner2020signatures}. These simulations on the timescales of tens of microseconds show transitions between LDL and HDL. 

Several molecular simulations have tried to illustrate the possibility of the existence of the two state picture also in room temperature water by examining the inherent structure of the liquid at zero-K\cite{accordino2011quantitative,wikfeldt2011spatially,pettersson2018two}. The idea of liquid water consisting of two co-existing liquids is invoked to rationalize various spectroscopic measurements of water\cite{myneni2002spectroscopic,wernet2004structure}. On the other hand, most molecular dynamics simulations of water at room temperature show that it is a homogeneous liquid\cite{characterizingdeboue,stirnemann2012communication} and that any heterogeneities in its structure arise from transient short-lived fluctuations\cite{henchman2016water,stirnemann2012communication,Kuhne2013}. At the heart of understanding this problem lies the question of the discovery of molecular probes of the local environments in a disordered liquid medium and subsequently, how to capture highly complex patterns in the hydrogen bond network on different length-scales.

Over the last three decades, since the advancement in the use of computer simulations, a wide plethora of different reaction coordinates or order parameters have been constructed to interrogate local environments in water. These include, to name a few, the tetrahedrality (q$_{tet}$)\cite{giovambattista2005structural,yan2007structure}, local-structure index (LSI)\cite{appignanesi2009evidence}, the distance of the 5th closest water molecule (d$_{5}$) to a central water\cite{cuthbertson2011mixturelike,tanaka2019revealing,saika2000computer} and local coordination defects \cite{gasparotto2016probing,agmon2012liquid,sciortino1989hydrogen,henchman2010topological}. Besides variables that quantify correlations in the water network, there have also been measures to quantify local density using variations of the Voronoi volumes which probes the free-space in the network and hence the density\cite{yeh1999orientational,stirnemann2012communication,bernal1959geometrical,ansari2018,ansari2019spontaneously,ansari2020}. All these various quantities are inspired by chemical intuition and are used to infer differences between ordered and disordered water environments. The vast majority of these quantities are designed using cutoffs in the number of water molecules, such as in $q_{tet}$ or $d_{5}$ or radially defined thresholds in the case of the LSI. If water is seen as a percolating directed liquid network with medium-to-long range correlations beyond the first solvation shell, the ability of these chemically inspired parameters to capture all the complexities of the water network, remains an open question.

In this work, we employ a series of state-of-the-art data science techniques to investigate the fluctuations underlying the free energy landscape of the TIP4P/2005  \cite{abascal2005general} model of liquid water at room temperature and also close to the critical point. Our strategy is implemented in three steps and lays the ground-work for a general framework through which fluctuations of water in different contexts may be studied. Firstly, we encode the information of water environments using a recently developed atomic-descriptor, the smooth-overlap of atomic positions (SOAP) which has the power of preserving rotational, permutational and translational invariances \cite{bartok2013representing}. This is used to compare water environments on different length scales and topologies to important milestones such as ice \cite{monserrat2020liquid}. 

In the second step, the dimension of the manifold in which the SOAP descriptors lie is estimated using the two nearest neighbors intrinsic dimension estimator (TWO-NN) \cite{facco2017estimating}. This quantity, known as the intrinsic dimension (ID), has been successfully used to characterize changes in the conformation of proteins\cite{jong2018data,sormani2019explicit} as well as phase transitions in simple classical and quantum Hamiltonians\cite{mendes2021unsupervised,PRXQuantum.2.030332}. The ID also feeds into the third and final step of our procedure in which the minima and transition states of the high-dimensional free energy landscape are located by using a density peaks clustering algorithm\cite{rodriguez2014clustering,rodriguez2018computing,d2018automatic}. 
%In the second step, distances obtained from SOAP which lie on a high-dimensional manifold, are used as input into a dimensionality-reduction algorithm, namely the two nearest neighbors intrinsic dimension estimator (TWO-NN)\cite{facco2017estimating}. 
%This procedure reveals the intrinsic dimension (ID) of the water network which is used in estimating the free energy \cite{rodriguez2018computing}. 
%Finally feeds into the final step which uses two advanced clustering procedures, namely density peak clustering (DPA) \cite{d2018automatic}. the uniform manifold approximation and projection (UMAP)\cite{mcinnes2018umap} to characterize the underlying free energy landscape.

%The free energy landscape of liquid water at room temperature as revealed by this procedure, illustrates that the underlying free energy landscape of water consists of one broad minimum. 
This procedure illustrates that, at room temperature, the free energy landscape associated with the local environment of water consists of one broad minimum that expands in several dimensions.
This feature is reproduced using both TIP4P/2005 as well as the many-body MB-pol potential\cite{mbpol1}. 
The fluctuations of the hydrogen bond network within this single minimum free energy landscape, however, occur in a high-dimensional space involving the coupling of many different degrees of solvent coordinates.
We find that most of the chemically inspired variables such as $q_{tet}$, LSI and d$_{5}$ do not adequately describe the environments of water in terms of their apparent LDL or HDL character. 
Instead, combining these variables with the SOAP-based descriptors provides a much more nuanced picture underlying the fluctuations. 
Specifically,  we show that the relevant variables needed to describe the fluctuations of water require both the oxygen and hydrogen atoms and that there are important hydrogen bonding interactions between the first and second solvation shell that affect the ID.
These features are not always captured by coordinates such as d$_{5}$ and LSI. 
We also explore how the SOAP descriptors evolve close to the critical point based on analysis of long trajectories in a recent work by Sciortino and Debenedetti\cite{debenedettiscience2020}. 
This also provides new insights into the complexity of the hydrogen-bond fluctuations in the supercooled regime.

The paper is organized as follows. We begin in Section 1 with a description of all the computational methods used. In Section 2, we report all on all our results where we discuss: our findings of the intrinsic dimensionality of the hydrogen bond network, the free energy landscape of liquid water at room temperature and the molecular origins of the associated high dimensional fluctuations. Within the results we also elucidate the behavior of the high dimensional fluctuations close to the critical point. Finally, we end with some conclusions and perspectives.

\newpage
\section{Methods}\label{methods}
\label{section:1}
\subsection{Molecular Dynamics Simulations}

All-atom molecular dynamics simulations (MD) of 1019 water molecules were performed using the
GROMACS 5.0 package\cite{bekker1993gromacs}. For most of the results we report in this paper, we use the TIP4P/2005\cite{abascal2005general} rigid water model. Besides this water model, we also compare some of our results with the MB-pol potential which is the most accurate in-silico potential reproducing many of the thermodynamic, dynamical and spectroscopic properties of water across the phase diagram\cite{babin2013development,reddy2016accuracy}. Energy minimization was first carried out to relax the system, followed by an NVT and NPT equilibration at 300K and 1 atmosphere for 10ns each. A timestep of 2fs was used for all the simulations. The NVT simulations were performed using the velocity-rescaling thermostat\cite{bussi2007canonical} with a time constant of 2ps, while the NPT runs were conducted using the Parrinello-Rahman\cite{parrinello1980crystal} barostat using a pressure coupling time constant of 2ps. The production run at 300K was carried out for 50 ns\cite{hockney1981computer} in the NPT ensemble.

Besides the simulations at 300K, we also analyzed molecular dynamics trajectories of supercooled water reported recently by Debenedetti and Sciortino which showed for the first time, that atomistic models such as TIP4P/2005 and TIP4P/ICE \cite{abascal2005potential} also display a second critical point. In these simulations, one observes fluctuations between high and low density phases of water. For more details on how these trajectories were generated, the reader is referred to the original manuscript\cite{debenedetti2020second}.

\subsection{Descriptors for Water Environments}
\begin{figure}[!htb]
    \centering
    \includegraphics[width=15cm]{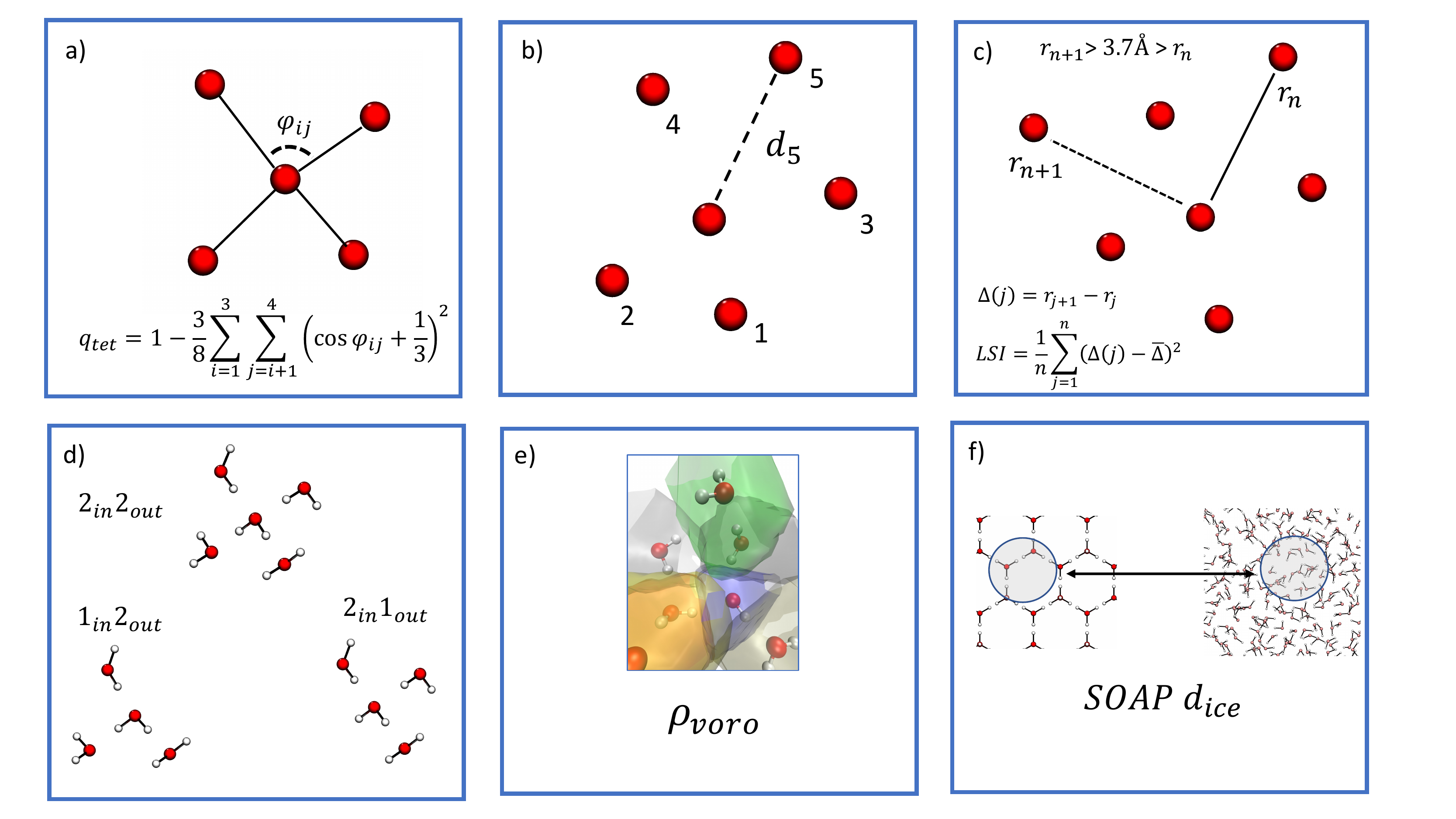}
    \caption{A visual schematic summarizing all the one-dimensional descriptors of water environments that were used in this work. a) $q_{tet}$, Tetrahedrality  associated with the four nearest neighbors. b) $d_5$, distance from the fifth nearest neighbor. c) $LSI$, Local Structure Index measuring the extent or order/disorder between 1st and 2nd solvation shells. d) Topological defects in water. Shown are some examples including $2_{in} 2_{out}$, $1_{in} 2_{out}$ and $2_{in} 1_{out}$ water molecules. The $in,out$ subscript correspond to the number of hydrogen bonds being accepted/donated respectively by the central water molecule e) $\rho_{voro}$, the local density estimated as the inverse of the Voronoi volumes corresponding to each water molecule. f) SOAP $ d_{ice}$ is the distance in the SOAP space from a given liquid water environment to the one present in hexagonal ice.}
    \label{fig:scheme1}
\end{figure}

As eluded to earlier, we use a combination of both chemical-intuition and data-science based descriptors to probe the local environment of water. While the former have been used to study the structure of water under different thermodynamic conditions\cite{tanaka2019revealing,russo2014understanding,characterizingdeboue}, the latter have been mainly developed for use in machine-learning applications\cite{ceriotti2019unsupervised,behler2007generalized}. In the following, we will detail the essential principles behind the descriptors used in this work (summarized in Figure \ref{fig:scheme1}).

\subsubsection{Chemical-Based Descriptors}

Tetrahedrality ($q_{tet}$, Fig. \ref{fig:scheme1}a ) measures the similarity between the first layer environment and a tetrahedron. It takes as input the angles computed taking as vertex the oxygen atom of the central water and all the possible couples generated by the four nearest neighbors, yielding a value of 1 for a perfectly tetrahedral environment\cite{chau1998new,lynden2005computational}. 
The $d_{5}$ parameter (Fig. \ref{fig:scheme1}b) is the distance from the fifth nearest neighbor to the central atom and reflects the extent of separation between the first and second solvation shells. A larger value of $d_{5}$ is interpreted as being a more open and locally ordered structure\cite{saika2000computer}. Fig. \ref{fig:scheme1}c) shows the functional form used to measure the LSI which is designed to probe the order in the first and second coordination shells. By examining all but one water molecule within a cutoff of $3.7$\AA{} the LSI distinguishes environments with a well separated first and second coordination shells, from those that are more disordered.
%variance of the neighbor shell dimension, where this dimension is defined by the difference of the distance from the central oxygen atom of two consecutive nearest neighbor oxygen atoms. The numbers of nearest neighbors considered in this definition are one more than those included within $3.7$~\AA, therefore including one atom outside the first solvation shell and probes how different are the distances inside the first shell from the inter-shell distances.
%are aimed at characterizing fluctuations that occur between the first and second hydration shell. 

All these variables do not explicitly include the hydrogen atoms. Numerous previous theoretical studies have shown that there are important correlations in the hydrogen-bond network created by the local topology which involves directed hydrogen bonds between water molecules\cite{henchman2010topological,gasparotto2016probing,policharge2020,ansari2020}. Figure \ref{fig:scheme1}  d) shows some examples of topological defects that can be created from the canonical $2_{in} 2_{out}$ (two hydrogen bond donor and two acceptor) water which we also examine in our work.

Although the previously described parameters are often interpreted in terms of high and low density environments, this can only be inferred indirectly. To obtain a more quantitative measure of density variations, we computed the Voronoi density ($\rho_{voro}$) as illustrated in Figure \ref{fig:scheme1} e).  $\rho_{voro}$ is computed as the inverse of the Voronoi-volume associated with a water molecule which is the sum of the volume of the oxygen and two hydrogen atoms \cite{yeh1999orientational,stirnemann2012communication,bernal1959geometrical}. 
%(i.e. the volume of the regions whose nearest atom are these atoms)
\subsubsection{Smooth Overlap of Atomic Positions (SOAP)}

Finally, Figure \ref{fig:scheme1} f) shows the last one dimensional descriptor that we used in this work, namely the SOAP distance from a hexagonal ice structure. This variable quantifies how different a local water environment in liquid water is from a water molecule obeying the ice-rules in the ice lattice. Depending on the choice of the various SOAP parameters, one can generate a wide variety of distance measures. Since the application of these types of descriptors in the study of condensed phase systems is quite new, we provide a more detailed discussion of the method next.

As indicated earlier, local atomic descriptors such as SOAP preserve important symmetries such as rotational, permutational and translational invariances when comparing different environments. The SOAP descriptor in particular, has successfully been applied in various contexts such as characterizing hydrogen bond networks\cite{appignanesi2009evidence} 
%\adu{References were for constructing neural network potential, and understanding local structure} 
in biological systems\cite{maksimov2021conformational,grant2020network,de2016comparing}, inorganic crystals\cite{bartok2018machine,reinhardt2020predicting} and also very recently in liquid water\cite{monserrat2020liquid}. 

Given a particular atomic species $Z_{i}$, one characterizes its local environment as a sum of Gaussian functions with variance $\sigma^{2}$ centered on each of the neighbors of a central atom including the central atom itself: 

\begin{equation}
\rho^{Z_{i}}(\mathbf{r})=\sum_{j} \exp \left ( \frac{-\left | \mathbf{r} - \mathbf{r_{ij}}\right |^2}{2\sigma ^2} \right)
\end{equation}

This atomic neighbour density can be expanded in a basis of radial basis functions and spherical harmonics as illustrated below:
\begin{equation}
   \rho^{Z_{i}}(\mathbf{r}) \approx \sum_{n=0}^{nmax} \sum_{l=0}^{lmax}\sum_{m=-l}^{l} c^{Z_{i}}_{nlm}(\mathbf{r}) g_{n}(r)Y_{lm}(\theta, \phi)
 \end{equation}
where the $c^{Z_{i}}_{nlm}(\mathbf{r})$ are the coefficients.
Given $\rho^{Z_{i}}(\mathbf{r})$ one can obtain the coefficients as

\begin{equation}
c^{Z_{i}}_{nlm}(\mathbf{r}) =\iiint_{\mathcal{R}^3}\mathrm{d}V g_{n}(r)Y_{lm}(\theta, \phi)\rho^{Z_{i}}(\mathbf{r}).
\end{equation}

The number of coefficients of the basis functions one chooses to compute, is bounded by the number of radial basis functions  $nmax$ and that of the angular basis functions $lmax$. The parameter $rcut$ identifies all molecules within some radial cutoff of the central atom. One can then define a rotationally invariant power spectrum as
\begin{equation}
     p(\mathbf{r})^{Z_1 Z_2}_{n n' l} = \pi \sqrt{\frac{8}{2l+1}}\sum_m c^{Z_1}_{n l m}(\mathbf{r})^*c^{Z_2}_{n' l m}(\mathbf{r})
\end{equation}

By accumulating the elements of the power spectrum into a vector of unit length $\mathbf{p}$, the distance between two environments $\chi$ and $\chi'$ is related to the SOAP kernel by the following expression

\begin{equation}
 d(\chi,\chi') ={1  -  K^\mathrm{SOAP}(\mathbf{p}, \mathbf{p'})} 
\end{equation}
 
where,

\begin{equation}
K^\mathrm{SOAP}(\mathbf{p}, \mathbf{p'}) = \left( \frac{\mathbf{p} \cdot \mathbf{p'}}{\sqrt{\mathbf{p} \cdot \mathbf{p}~\mathbf{p'} \cdot \mathbf{p'}}}\right)
\end{equation}

For our simulations of bulk water, the SOAP descriptor for a water molecule is constructed involving different combinations of the oxygen and hydrogen atoms and their environments. The first descriptor ($\vec{\textbf{O}}$), is formed by computing the power spectrum of the density constructed by placing Gaussian functions on only the oxygen atoms within a certain radius centered about the position of the oxygen of the central water molecule. The other two descriptors include the hydrogen atoms of the water molecule. Since a water molecule contains two hydrogen atoms, it is necessary to choose the centers in a way that would make new descriptor invariant to the permutation the two indices. This was achieved by averaging the power spectra generated with centers on each of the hydrogen atoms ($\vec{\textbf{H}}_{ave}$). In order to preserve information about the possible asymmetries present in the environment of the two hydrogen atoms, the absolute value of the difference in the descriptors was also considered ($\vec{\textbf{H}}_{dif}$). Using these soap descriptors, we focused the ensuing analysis on three variations: $\vec{\textbf{O}}$, ($\vec{\textbf{O}}$, $\vec{\textbf{H}}_{ave}$) and finally, ($\vec{\textbf{O}}$, $\vec{\textbf{H}}_{ave}$,$\vec{\textbf{H}}_{dif}$).

%\alex{I'm not able to find where we state the frequency of sampling of the frames, it could be introduced here.}

The SOAP descriptors were constructed using the Dscribe package\cite{himanen2020dscribe}. In practice 10 randomly chosen water molecules were selected from each frame with a sampling frequency of 4ps. This was done to ensure independence of points by reducing the effects of spatial and temporal correlations between sampled environments.  In total, $120000$ points were extracted for a radial cutoff of 3.7 and 6.0 \AA. These cutoffs were chosen to enclose the first and second hydration shell of water respectively. Our analysis was done using 8 ($\bf{nmax}$) radial and 6 ($\bf{lmax}$) angular basis functions. These values are similar to those used in previous studies\cite{monserrat2020liquid,bartok2018machine}. With the power spectrum of the SOAP-based environments in hand, one can compute distances between the different local environments in water and other milestone structures for example ice. In this work, for most of our analysis, we focus on comparing the local environments in water to those in hexagonal ice although. In the rest of the manuscript, this distance is referred to as $d_{ice}$. One can also select many different milestone structures to compare with. For example, Petersson and co-workers have recently proposed the possibility of low-density liquid environments arising from fused dodecahedron structures\cite{pettersson2019}. SOAP distances can then be constructed with respect to this structure. When this is done, we will make reference to this distance as $d_{dod}$. Bingqing et al. and Pavan et al. have recently used SOAP based descriptors to compare environments in water to different phases of ice by building an average SOAP kernel\cite{monserrat2020liquid,capelli2021ephemeral} which allows for examining changes in the global structure. However, it should be stressed that, while building average descriptors is sufficient for describing the global properties of phases, averaging over all the environments, washes out important details on the local molecular structure.

\subsection{Data Science Protocol}

Extracting the SOAP descriptors as outlined previously, results in high dimensional power spectra. Specifically, the dimensions for the three SOAP descriptors outlined earlier, are 252, 504 and 2856 for $\vec{\textbf{O}}$, ($\vec{\textbf{O}}$, $\vec{\textbf{H}}_{ave}$) and ($\vec{\textbf{O}}, \vec{\textbf{H}}_{av} \vec{\textbf{H}}_{dif}$) respectively. The high dimensionality of these spectra implies the need of using advanced techniques to extract the meaningful information. In the following, we review the data mining techniques that we employed.

%The amount of data, and its high dimensionality, implies the use of advanced techniques to deal with all this information. 
\subsubsection{Intrinsic Dimensionality (ID)}

In high dimensional data, the existence of correlations between the different variables describing each data point implies that the system of interest likely resides in a lower dimensionality landscape. Take for example, a set of points in three dimensions - if they are distributed randomly, the ID would be three. However, correlations between the coordinates could result in the data points lying only on the surface of the sphere which would in turn lead to an ID of 2.  Computing the ID is closely related with dimensionality reduction techniques \cite{mika1998kernel,jolliffe2005principal,kruskal1978multidimensional}, in which the data set is projected into a lower dimensional space for data analysis, visualization and interpretation. The ID represents the minimum dimensionality in which these techniques can be employed without an important information loss. A correct knowledge of the ID should inform the choice of the space one examines the fluctuations of the system. In the case of our study, the ID plays an important role in estimating a point dependent density function, which ultimately affects the extraction of the free energy as outlined later. 

%The methods for estimating the intrinsic dimension fall broadly into three categories; fractal methods, projection methods and nearest neighbor methods. Projection methods search for a subspace to project the data in by minimizing a projection error. Fractal methods estimate the ID by counting the observed points in a neighborhood around point and estimating how it scales with increasing distance. Nearest-Neighbors methods which assume local uniformity infer the ID based on statistics of nearest neighbor distances\cite{pettis1979intrinsic,verveer1995evaluation,levina2005maximum}.
In this work, we made use of a recently developed technique namely the Two-NN estimator\cite{facco2017estimating}, which estimates the ID based on information of the first and second nearest neighbor of data points. The method has been successfully applied in studying different molecular systems\cite{ansari2019spontaneously,carli2020candidate,jong2018data}. The main ideas are sketched below. Assuming local uniformity of the data set, it can be shown the ratio of second nearest neighbor and first nearest neighbor distances ($\mu = r_2 /r_1 $) yields the following distribution:

\begin{equation}
    P( \mu ) = \frac{d}{\mu ^{d +1}}
\end{equation}

where d is the ID. Assuming all sampled ratios $\mu_{i}$'s are independent, it can be shown that the ID can be estimated by the following formula 

\begin{equation}
    d = \frac{N}{\sum_{i}{ \log(\mu_i})}
\end{equation}

Using the SOAP distances, we estimated the ID of the environment around a water molecule. The physical significance of the ID is that it corresponds to the minimum number of independent order parameters or reaction coordinates required to describe, in our case, the environment around a water molecule. In this way, one can also quantify using the ID, the amount of information that is gained or lost when including different variables\cite{camastra2016intrinsic}.
%(This sentence is needed since allow us to compare the amount of info by comparing $I_d$s)}

\subsubsection{High Dimensional Free Energy and Clustering} 

Constructing the free energy landscape of water requires an understanding of the relevant variables that characterize its underlying structural fluctuations. 
One of the common strategies in this context is to look at the probability densities along variables such as those introduced earlier like $q_{tet}$, LSI and $d_{5}$\cite{cuthbertson2011mixturelike,appignanesi2009evidence,wikfeldt2011spatially}. 
However, this implicitly assumes that there is no information loss when this projection is done and furthermore, that the variable correctly encodes for the physical or chemical process that one is interested in. 
In the last two decades, there has been a growing number of advanced techniques developed to automatically identify important degrees of freedom\cite{tenenbaum2000global,roweis2000nonlinear} and construct free energies in high dimensions\cite{carreira2000mode,gasparotto2014recognizing,gasparotto2018recognizing,rodriguez2018computing,geissler1999kinetic}. For a more elaborate discussion of these techniques, the interested reader is referred to a recent review on the topic\cite{glielmo2021unsupervised}.

In this work, we employ a recently developed Point Adaptive K-nearest estimator(PAK)\cite{rodriguez2018computing} that avoids the need of any projection and  has been used to study a wide variety of complex molecular systems\cite{carli2020candidate,jong2018data,sormani2019explicit}. In brief, the method uses the ID as a parameter to construct a point dependent density ($ \rho_i $). 
This density is computed by adding a linear correction to the standard $k$-nearest Neighbor estimator in which the density is computed as $\rho_i = \nicefrac{k_{i}}{r_{k_{i}}^{d}}$ and the $k_i$'s are chosen in order to minimize the errors in density. The point dependent free energy is then taken to be $-\Log( \rho_{i})$. Previous work has shown that this methodology works very well at providing accurate estimates of the errors in free energy up to dimensions as large as 8\cite{rodriguez2018computing}.

With the point dependent free energies in hand, the independent minima in the free energy landscape (the clusters) are determined using a form of the modified density peak clustering algorithm (DPA)\cite{d2018automatic}. This is an unsupervised extension of the original density peak clustering\cite{rodriguez2014clustering} which automatically determines the clusters keeping only those that are statistically significant using a confidence interval $z$. Unless otherwise stated the z-value chosen for most of the analysis presented is set to 2.5.

Finally, the results of PAK and DPA have been visualized and interpreted with the help of the uniform manifold approximation and projection (UMAP)\cite{mcinnes2018umap}. UMAP provides a convenient way of visualizing the high dimensional free energy in two dimensions as has been done in several recent applications\cite{Becht2019}.

\section{Results and Discussion}

\subsection{ID analysis of Hydrogen-Bond Network}

%One of the principles that is emerging from the analysis of data coming from atomistic molecular dynamics simulations of biological systems, is that fluctuations involve several degrees of freedom within a rather high dimensional manifold \cite{ceriotti2019unsupervised,glielmo2021unsupervised}. 

%The dimension of this manifold, it is, the minimal number of variables to describe these fluctuations, can be estimated directly from the simulations. The Two-NN estimator, that we apply in this work, has been successfully used to characterize changes in the conformation of proteins\cite{sormani2019explicit,jong2018data} as well as phase transitions in simple classical and quantum Hamiltonians\cite{mendes2021unsupervised,PRXQuantum.2.030332}. Therefore, in the case of the water environments, it should reflect the spatial correlations involving involving both the oxygen and hydrogen atoms across the different length scales considered.   

Using the Two-NN estimator, we extracted the ID of the hydrogen bond network using the SOAP descriptors described earlier. In the context of water fluctuations, the ID provides a quantitative measure of the changes in the information content on adding several features when describing the local water environment. Specifically, in our analysis, we systematically examine how the ID changes when increasing the cutoff from layer 3.7 \AA{} to the second 6.0 \AA{} and when adding hydrogen atoms to the descriptors.

\begin{figure}[!htb]
    \centering
    \includegraphics[width=15cm]{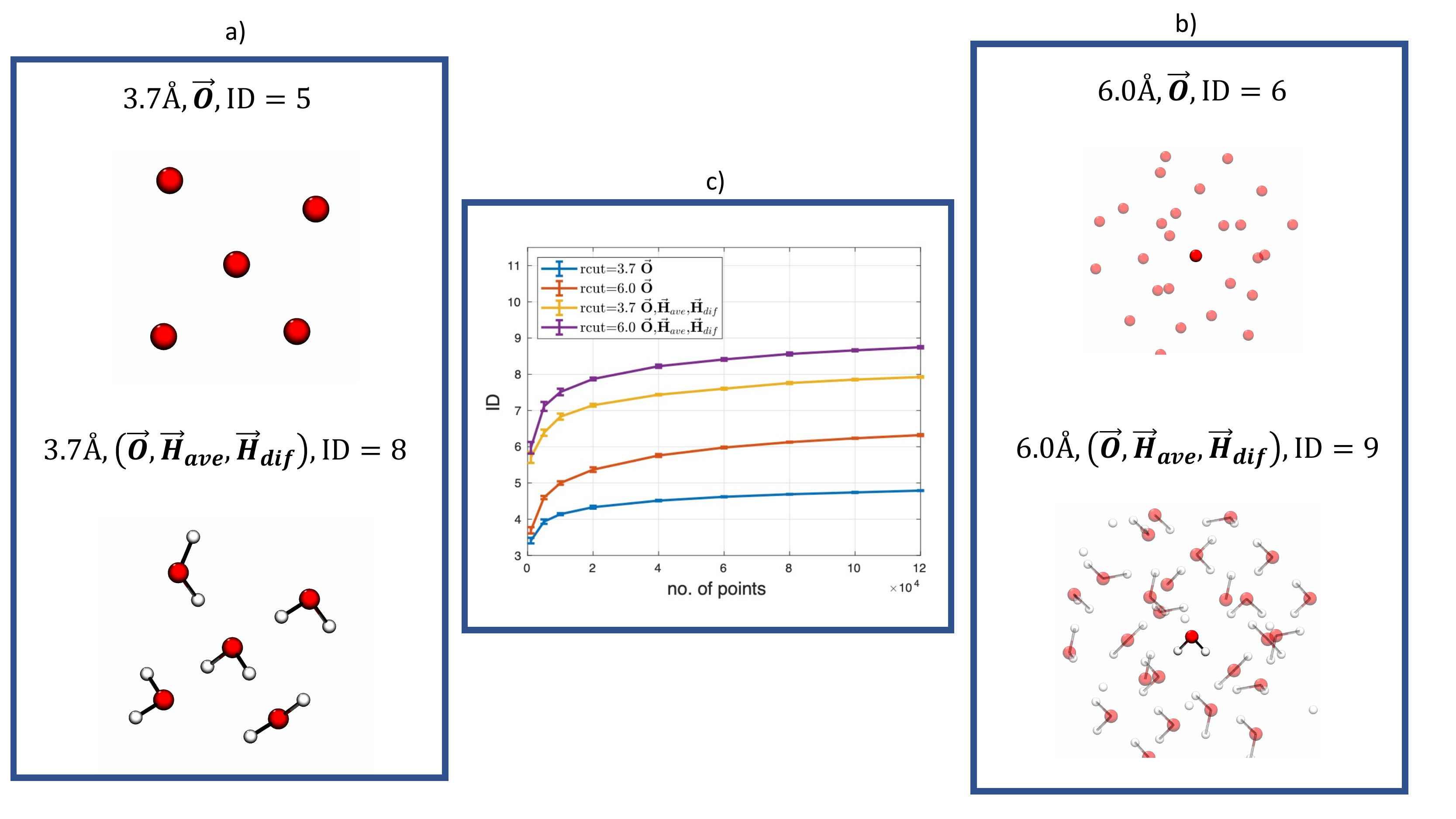}
    \caption{Panel a)  shows 3.7 \AA{} environment involving only oxygen atoms (top) and the same environment with hydrogen atoms (bottom). The intrinsic dimensionality is found to increase by 3 when hydrogens are included. b) Shows 6.0 \AA{} environment involving only oxygen atoms (top) and the same environment with hydrogen atoms (below). The intrinsic dimensionality is found to increase by 3 when hydrogens are included.  c) Shows the scaling of the intrinsic dimensionality with the number of points of the data set. }
    \label{fig:scheme}
\end{figure}

Figure \ref{fig:scheme} summarizes the results obtained from the ID analysis. Panels a) and b) schematically illustrate the local environments that are included with the corresponding inferred IDs. Panel c) shows the convergence of the ID as a function of the number of data points. Essentially, the ID can be increased in two possible ways: firstly by including or excluding the chemical species in a water molecule namely oxygen or hydrogen atoms, and secondly, by increasing the size of the solvation shell of the local water environment.

Interestingly, the ID analysis shows there is a much bigger change in the importance of including hydrogen atoms into the descriptors compared to expanding the radial cutoff. For both the 3.7\AA{} and 6\AA{} radial cutoffs, the ID increases by 3 upon the inclusion of the hydrogen atoms. On the other hand, moving from the smaller to larger radial cutoff increases the ID by a unit value.  
Several of the chemical-based order parameters described earlier in Figure \ref{fig:scheme}, for example, $q_{tet}$, $d_{5}$ and the LSI do not explicitly include the hydrogen atoms and therefore are very likely to miss out important coordinates needed to characterize water environments. 

To understand better the molecular origins of these differences in the ID, one can examine the effect of the inclusion of the hydrogen atoms when comparing different water environments.
For instance, we can take the two defects that are shown in Figure \ref{fig:scheme1} panel d namely, $1_{in}2_{out}$ and $2_{in}1_{out}$ and compute the distribution of the SOAP distances within each class of defect type and between the two defects. In SI Figure \ref{fig:si_inter_intra} it can be seen that when using only the oxygen atoms ($\vec{\textbf{O}}$), the distributions within and across different defects are almost identical. However, upon adding the hydrogen atoms contributions ($\vec{\textbf{O}}$, $\vec{\textbf{H}}_{ave}$,$\vec{\textbf{H}}_{dif}$), the distributions are different with a slight bias towards higher values in the case of the inter-group distance distributions. Although there is clearly significant overlap in all these distributions, our analysis shows that by only using ($\vec{\textbf{O}}$), there is important information about the hydrogen bond network that is lost.

The changes in the ID has important implications on our understanding of the free energy landscape of water. Firstly, the hydrogen bonds between water molecules involve directed dipole-dipole interactions which arise from the asymmetry in the position of the hydrogen atoms. This feature of the chemistry is clearly reflected in the change of the ID upon including the hydrogen atoms. At the same time, the presence of medium-to-longer range structural and orientational correlations in the network are manifested also in the increase in the ID albeit to a smaller extent than the role of directionality\cite{cuigalli2014}. In some sense, the enhancement in the ID from the oriented hydrogen bonds within 3.7\AA{} is strongly coupled to the order or disorder at longer distances. 

In addition to the effect of the ID on the combination of the variables shown in Figure \ref{fig:scheme}, we also examined the effect of the two hydrogen atom based SOAP descriptors namely $\vec{\textbf{H}}_{ave}$ and $\vec{\textbf{H}}_{dif}$. For both the 3.7\AA{} and 6.0\AA{} environments, eliminating $\vec{\textbf{H}}_{dif}$ reduces the ID by one unit to 7 and 8 respectively. This effect on the ID, indicates that there are important asymmetries involving the environments of the two hydrogen atoms that can donate hydrogen bonds. A clear example of this, is shown in Figure \ref{fig:scheme1} d) illustrating the creation of different types of topological defects.

\subsection{Free Energy Landscape of Liquid Water}

Having computed the ID, we are now in a position to generate the high-dimensional free energies using the PAK-Nearest density estimator. One of the challenges in constructing the point-free energies is that the error grows with the ID\cite{rodriguez2018computing}. We thus begin by focusing our analysis on using the SOAP descriptor environments consisting of only the oxygen atoms with a radial cutoff of 3.7\AA{}. 

%\ali{Please add citation(s)}

Visualizing the high-dimensional free energies is extremely challenging. With the PAK free energies, we performed the modified density peak clustering using a confidence interval of z=2.5 which indicates the presence of one big cluster. The presence of one cluster at this z value is reproduced across several different water molecule environments suggesting that this is not an artifact of statistical fluctuations.% To provide further credence to the clustering result, we validated our result with an additional density based clustering method namely the Hierarchical Density-Based Spatial Clustering of Applications with Noise (HDBSCAN)\cite{campello2013density,campello2015hierarchical}. The HDBSCAN method\cite{McInnes2017}, provides a consistent result when directly applied to the SOAP descriptors namely that a single big cluster is obtained (see the condensed tree plot in Figure \ref{fig:si_hdbscan}).  

\begin{figure}[!htb]
    \centering
    \includegraphics[width=\textwidth]{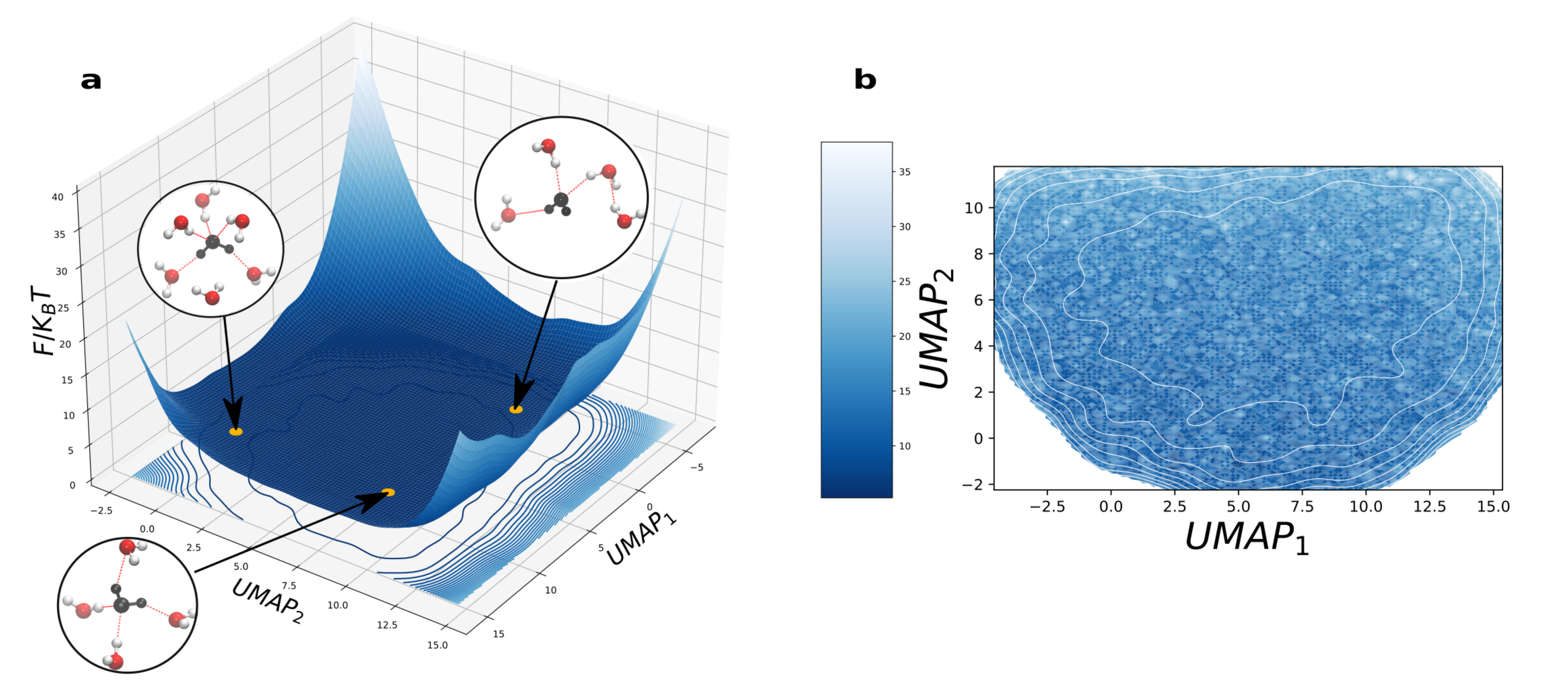}
    \caption{a)Free energy surface constructed in 2D UMAP manifold reveals a single basin. Also shown are three different water environments which are separated by no significant barriers indicating that fluctuations between different structures occurs within a flat free energy landscape.  b) Contour plot of the 2D UMAP manifold colored by the actual free energy values is also consistent with a rough but rather flat basin.}
    \label{fig:free3d_ice}
\end{figure}

In order to gain a more visual inspection of the free energy landscape, we project the SOAP coordinates in two dimensions using the UMAP method\cite{mcinnes2018umap}. Although reducing the dimensionality below the ID leads to an unavoidable information loss, the UMAP method has shown to fairly preserving the global structure of the data\cite{diaz2019umap} providing a convenient way to visualize the free energies. The left panel of Figure~\ref{fig:free3d_ice} shows the free energy surface obtained with UMAP. Interestingly, the landscape is characterized by a very broad and rather flat free energy with small barriers ($k_{B}T$) separating shallow minima. These minima are characterized by water environments that are quite diverse as seen in the three snapshots taken at various points in the basin. These fluctuations between defective and non-defective environments without deep minima, is consistent with the presence of short-lived (between fs-ps) heterogeneities in water\cite{stirnemann2012communication,henchman2010topological}. The right panel of Figure~\ref{fig:free3d_ice} shows a 2d-contour map along the UMAP coordinates colored by the free energy which more clearly illustrates these features and confirms that at these temperatures, liquid water is indeed a homogeneous liquid\cite{characterizingdeboue,soper2019water}.  
The UMAP manifold for two other datasets corresponding to different choices of water molecule environments, were found to be essentially the same suggesting the main features of the landscape are not artefacts of statistical fluctuations (see SI Figure \ref{fig:si_umap_3}).

The preceding analysis is performed on the SOAP descriptors involving only oxygen atoms within 3.7\AA{}. Since the ID changes quite significantly when including the hydrogen atoms we repeated the PAK and UMAP analysis with the other SOAP variable combinations. Expanding the solvation environment does not lead to any significant changes in the free energy landscape. More quantitatively, Figure \ref{fig:si_free_error} in the SI shows a scatter plot of the free energies comparing the results using 
$\vec{\textbf{O}}$, and ($\vec{\textbf{O}}, \vec{\textbf{H}}_{av} \vec{\textbf{H}}_{dif}$) both with a radial cutoff 3.7\AA{} using the same water coordinates. The two free energies are very well correlated with each other and shows that while the hydrogen atoms expand the dimensionality of the free energy landscape, the key physical features remain very similar. As pointed out earlier, the full descriptor including the hydrogens is important for distinguishing different defect environments and therefore, the effects on the underlying free energy may become more pronounced in regimes where these defects are enhanced such as the air-water interface\cite{policharge2020}.

The current results have been extracted using the TIP4P/2005 water model. To validate our results, we repeated our procedure of extracting the ID, PAK and UMAP projection on trajectories of the MB-pol water model at room temperature\cite{reddy2016accuracy}. MB-pol is a many-body interaction potential which is the most accurate in-silico model for neutral and non-dissociative water across the phase diagram\cite{mbpol1,reddy2016accuracy}. In this model, we also observe the same features, namely the presence of a broad and flat free energy landscape (see SI Figure \ref{fig:si_umap_mbpol}). 

%In order to understand the free energy at 300K, the PAK-Nearest density estimator was applied to soap descriptor environments consisting of only oxygens at a radial-cutoff of 3.7 $\AA$. The main motivation being the small errors compared to when Hydrogen atoms are included see  figure \ref{fig:free_error}.Furthermore the free energy of the same environments constructed with these two variations reveals a linear relation with a correlation coefficient of 0.707 with a mean squared deviation of ~ $ 2 k_{B} T$ see SI figure \ref{fig:free_error}.

%\alex{In Fig. \ref{fig:free3d_ice}, the contour map should be bigger and colored by free energy.}

%The 2d projection of the three datasets in the umap manifold colored by the distance from ice is shown in SI figure \ref{}(fig:$umap_ice$) . The similar values of reveals $d_{ice}$ suggests we are sampling the same part of the free energy landscape.  

%Using the umap manifolds as a guide, the modified density peak clustering was carried out varying the only free parameter in our analysis as described in the previous section. The smallest valuesof z that yielded consistent clusters across the three datasets was found to be at z= 2.5. This is consistent with only a single cluster. The centers, of the datasets were found to be different across the dataset which is consistent with a free energy basin. 
 
%This result was crosschecked by using a k-peaks clustering developed in order to detect basins and yielded one cluster in the range of z=0.6 to 1.0 .
%Similar analysis was carried in the range of temperatures from 230-300 K was found to be consistent with a basin. 
 
\subsection{Molecular Origins of High Dimensional Fluctuations}

%\alex{Since here we separate by defects, it would be interesting to check what happens with the inclusion of Hydrogen atoms when computing the SOAP descriptors.}

The preceding analysis of the ID shows that the fluctuations involving the hydrogen bond network involve a rather larger number of solvent degrees of freedom moving in different directions. In the following, we will examine the correlations that exist between the various chemically inspired coordinates such as $q_{tet}$, $d_{5}$, LSI and Voronoi density ($\rho_{vor}$), as well as the new SOAP-based descriptor, $d_{ice}$ that was described earlier. Note that in this analysis a radial cutoff of 3.7 \AA{} is used. Also in this discussion we restrict our analysis to results in which $d_{ice}$ was computed with only oxygen atoms. It is worth noting, that the trends in behavior were found to be similar when the hydrogen atoms were included as well as when the radial cutoff was increased to 6.0 \AA{} (see SI Figure \ref{fig:si_iceh} and SI Figure \ref{fig:si_ice6}).

\subsubsection{Tetrahedrality and $d_{ice}$}

\begin{figure}[!htb]
    \centering
    \includegraphics[width=15cm]{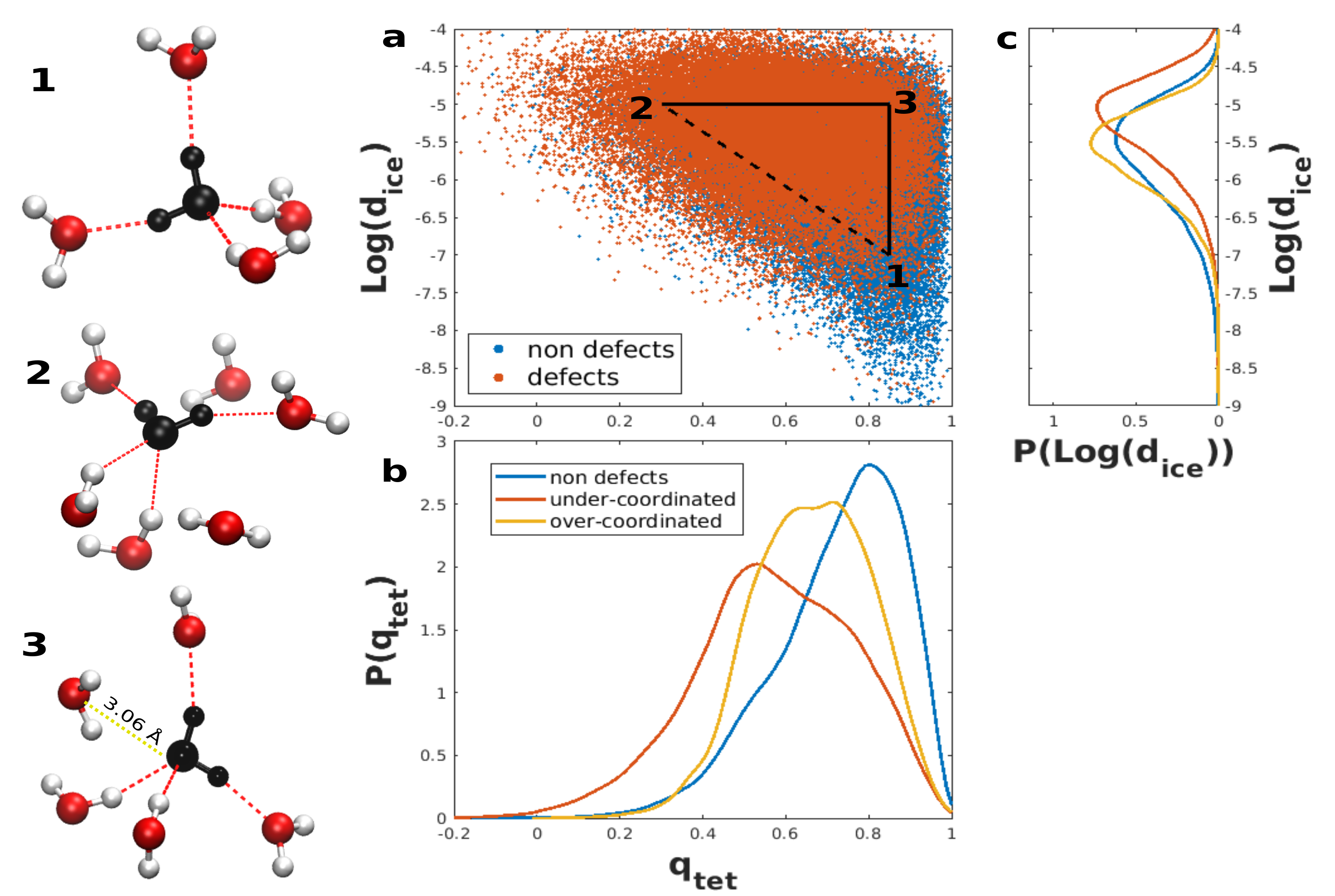}
    \caption{Figure  a) shows the scatter plot of $q_{tet}$ versus Log($d_{ice}$) with 3 points which correspond to the environments shown on the left. The points are colored to differentiate between defects and non defects. A large overlap is found between defects and non defects. Panels b) shows the probability density distributions along $q_{tet}$  and Log($d_{ice}$) for non-defects, and, under and over coordinated defects as defined in the main text.}
    \label{fig:tetrahedrality}
\end{figure}

We begin by showing in Figure \ref{fig:tetrahedrality} a scatter plot of $q_{tet}$ and $\Log(d_{ice})$. Also highlighted are 3 points (1-3) which are illustrated in the panels to the left of the Figure. The $q_{tet}$ and Log($d_{ice}$) values refer to that of the central water colored in black. The dashed line connecting points 1 and 2 corresponds to the regime where these two variables are strongly correlated with each other. Specifically, point 1 is a water environment that has a high tetrahedrality and a large negative value of $\Log(d_{ice}$) implying that it is closer in distance to a locally-ice like environment. Point 2 on the other hand has a low tetrahedrality and is non ice-like. The origin of the difference between these two structures is seen more clearly in the bottom panels where we observe that in point 2, there is an asymmetry in the angles between the donating and accepting side that are used to compute the tetrahedrality.

Perhaps the more surprising aspect of Figure \ref{fig:tetrahedrality} are all the points that lie above the dashed line. One of these limiting cases is shown by point 3 which has a high tetrahedrality but a large distance from ice using the SOAP coordinates. This environment illustrated in the bottom panel of Figure \ref{fig:tetrahedrality} (Point 3) shows that there is a water molecule that is within the first-shell ($\sim$ 3.23 \AA\ ) that is not hydrogen bonded to the central water. Nonetheless, the angles that are used to extract $q_{tet}$ involving only the nearest four neighbours do not include this water molecule and thus the central water is flagged incorrectly as a tetrahedral environment. 

Also shown in Figure \ref{fig:tetrahedrality} are the overlapped scatter plots for the defective and non-defect water molecules. Recall that defect waters are those which break the ice rules of accepting and donating 2 hydrogen bonds. Figure \ref{fig:tetrahedrality} b) and c) show the 1-d distributions for $\Log(d_{ice})$ and $q_{tet}$ for non-defects, under-coordinated (defined as when the sum of the number of donating and accepting hydrogen bonds is less than 4) and over-coordinated defects (where the sum of the number of donating and accepting hydrogen bonds is greater than 4). In this case, we see that while $q_{tet}$ and $\Log(d_{ice})$ are both characterized by differences in their average values for defect and non-defect populations, the former appears to show larger variation across the different environments. 

\subsubsection{$d_{5}$ and $d_{ice}$}

\begin{figure}[!htb]
    \centering
    \includegraphics[width=15cm]{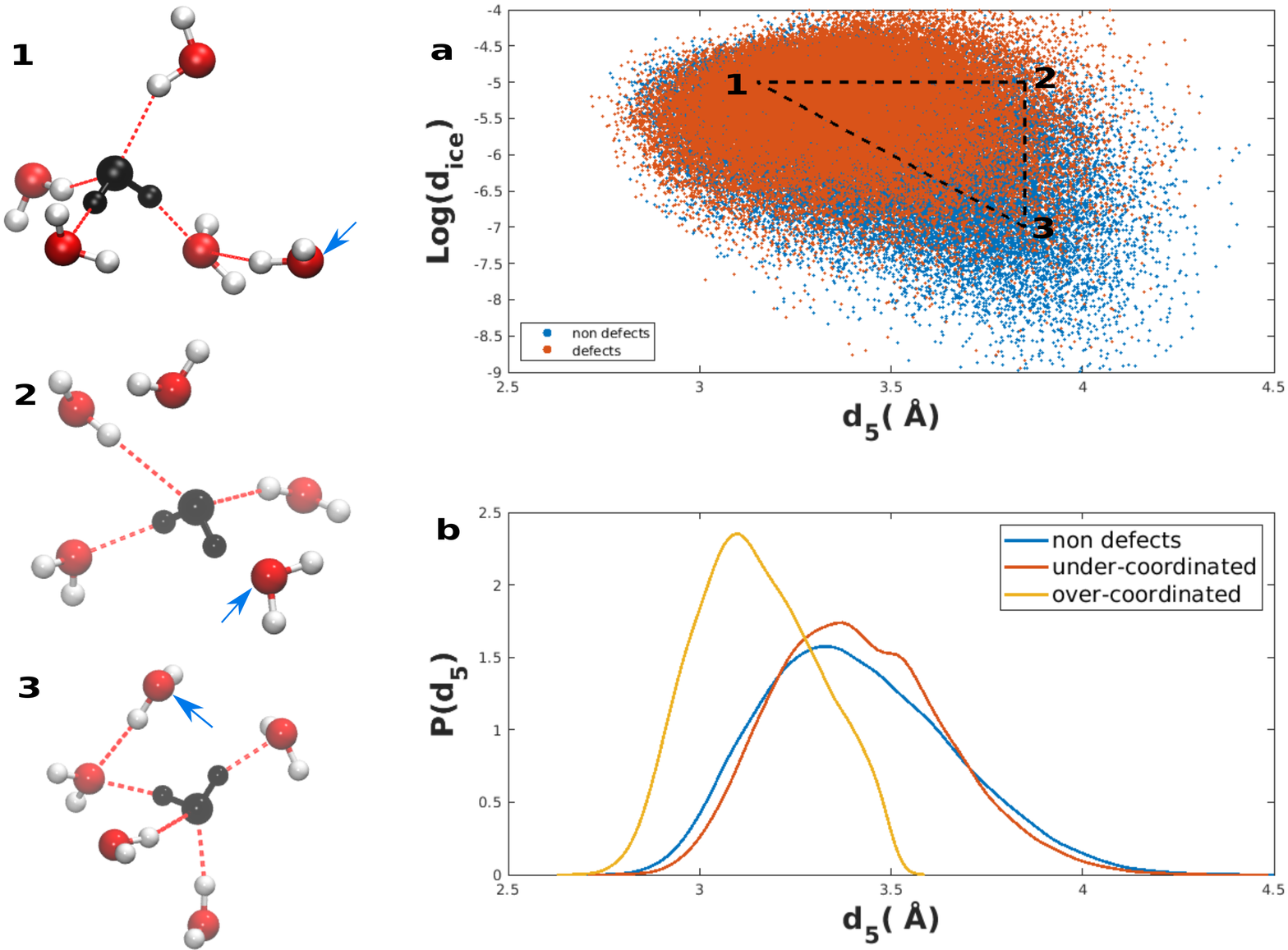}
    \caption{Panel a) shows the scatter plot of $d_{5}$ versus Log($d_{ice}$) with the 3 numbered configurations corresponding to the environments shown pictorially on the left. The blue arrow points to the water molecule that satisfies the $d_{5}$th criterion. Panels b and c show the probability densities obtained along $d_{ice}$ and $d_{5}$ for the defective and non-defective water molecules.}
    \label{fig:d5}
\end{figure}

Figure  \ref{fig:d5} shows the analysis performed on $d_{5}$ and $\Log(d_{ice})$. The $d_{5}$ parameter was designed in order to quantify fluctuations that occur between the first and second hydration shell\cite{cuthbertson2011mixturelike}. Specifically, a larger $d_{5}$ has been interpreted as a water environment that is more open and low-density like, while smaller values of $d_{5}$ as compact and high-density like. 

Similar to that analysis, we illustrate three landmark points in the scatter plot which are illustrated to the left of Figure \ref{fig:d5}. The water molecules referred to by the blue arrow correspond to waters that satisfy the $d_{5}$th criterion. The fluctuations along the line connecting points 1 and 3 reflect changes where the two parameters are well correlated: point 1 is a locally tetrahedral environment where the $d_{5}$ water resides in the second shell and separated by two hydrogen bonds from the central water, while in point 3, the $d_{5}$ water undergoes a large fluctuation bringing it from 3.7 \AA{} to within 3.1 \AA{} of the central water.

The fluctuations along the points 1-2 and 2-3 are more non-trivial as it shows that both $d_{ice}$ and $d_{5}$ play an important role in characterizing the local environments independently. Although point 2 has a high $d_{5}$ of approximately 3.7 \AA{}, asymmetries in the hydrogen bonds between the donating and accepting side of the first shell, renders it with a local configuration that is non-ice like. Examining the constrained distributions of the $d_{5}$ for the non-defects and under/over coordinated defects as before, shows that unlike $q_{tet}$, $d_{5}$ is much less sensitive in distinguishing these different environments. Intuitively, this is because the $q_{tet}$ is a parameter that uses the 4 nearest neighbours while $d_{5}$ focuses on just a single water molecule that fluctuates between the first and second hydration shell.

\subsubsection{Voronoi density$( \rho)$ and $d_{ice}$} 

\begin{figure}[!htb]
    \centering
    \includegraphics[width=15cm]{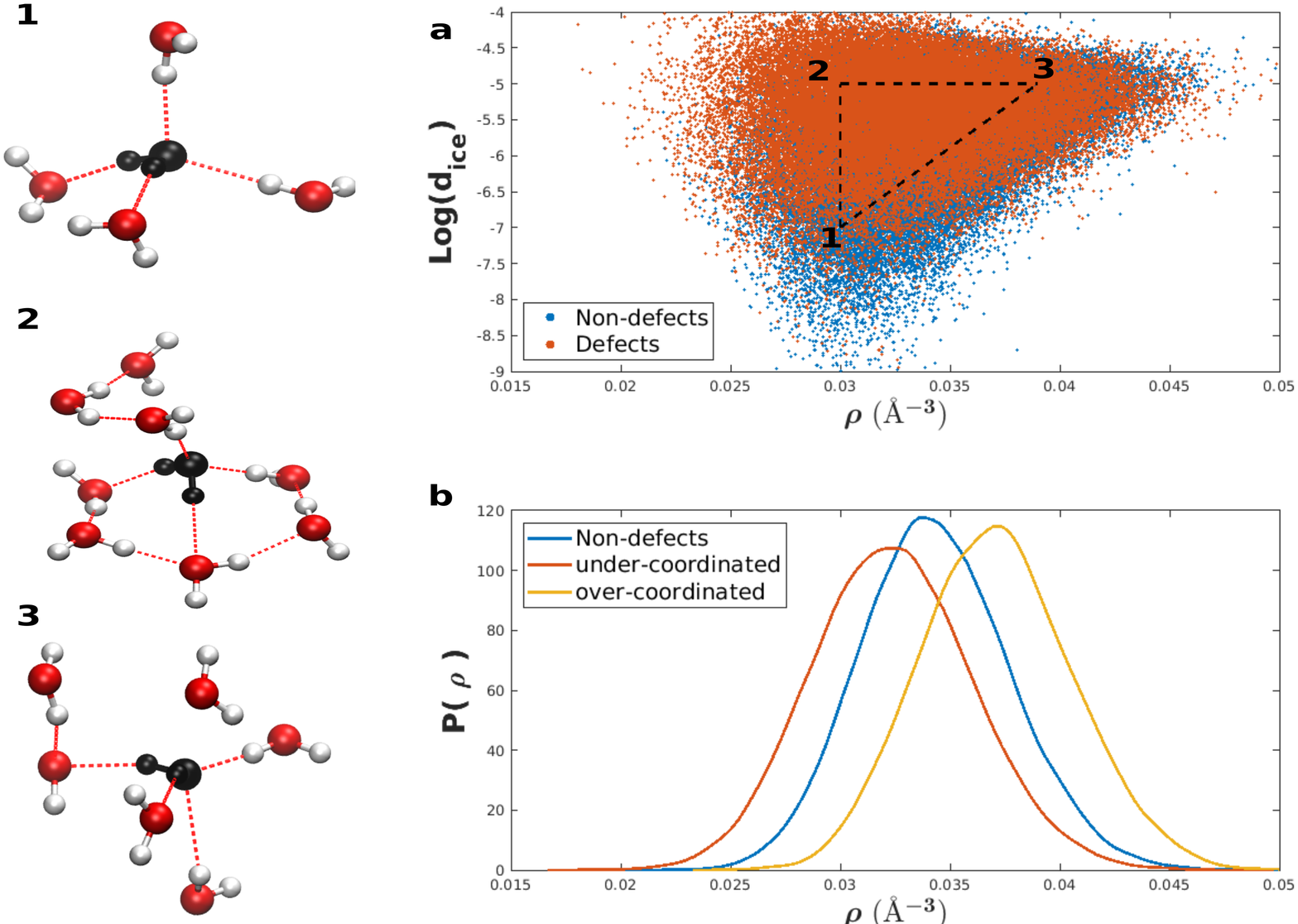}
    \caption{Panel a) shows the scatter plot of $\rho$ versus Log($d_{ice}$) with the 3 numbered configurations corresponding to the environments shown pictorially on the left. Panels b and c show the probability densities obtained along $d_{ice}$ and $\rho$ for the defective and non-defective water molecules.}
    \label{fig:voronoi}
\end{figure}

Figure \ref{fig:voronoi} shows the analysis performed on the Voronoi density $\rho $ and $\Log(d_{ice})$. The $\rho$ parameter has been used previously in the literature \cite{yeh1999orientational,stirnemann2012communication,bernal1959geometrical} to quantify local density fluctuations in liquid water at different thermodynamic conditions. As done in the previous analysis, a series of landmark points are illustrated to aid with the discussion. The fluctuations along the line connecting points 1 and 2 reflect changes where the two parameters are well correlated: point 1 is a low density open and ice-like environment, while point 2 corresponds to a higher density environment with 8 neighboring waters within 3.7 \AA. As expected, this high density fluctuation leads to the creation of an environment with a low value of $\Log{(d_{ice})}$.

Moving along points 1-3 and 2-3 confirms again, the importance of understanding the fluctuations of the network using a combination of several different variables. Point 2 is a high density environment created by a water molecule that participates in a six-membered ring and maintains a local tetrahedral order and therefore has a low value of Log($d_{ice})$. Point 3 on the other hand, corresponds to a low density environment but the orientations of the nearby water molecules do not have a local tetrahedral structure therefore leading to a higher value of Log($d_{ice})$. Figure \ref{fig:voronoi} c) shows the $\rho$ parameter performs rather well compared to the $d_{5}$ at distinguishing over and undercoordinated defect water molecules. It is interesting to note however, that there significant overlap in the densities for water molecules that accept and donate 2 hydrogen bonds and both under/over coordinated waters.

\subsubsection{LSI and $d_{ice}$}

Finally, we conclude this section with a comparison of the LSI and SOAP based parameter $d_{ice}$. The LSI variable was designed in order to quantify fluctuations between more ordered and disordered environments due to fluctuations at the boundary between the first and second solvation shell \cite{shiratani1998molecular,appignanesi2009evidence}. Specifically, a larger value of LSI has been interpreted as a water environment that is more open and characterized by a separation between first and second shell while smaller values of LSI, correspond to environments without a well separated first and second shell owing to the presence of interstitial waters. 

The dashed line connecting points 1 and 3 corresponds to the regime where these two variables are well correlated with each other. Specifically, point 1 is a water environment that has a large LSI and a large negative value of Log($d_{ice}$) implying that it looks locally like an ice-like environment. Point 3 on the other hand has a low LSI and is non-ice like. The origins of the difference between these two structures is seen more clearly in the leftmost panels where we observe in point 3 the presence of several interstitial water molecules.

The region connecting the points 2 and 3 illustrates the challenge in interpreting the LSI coordinate in terms of the local order/disorder.  Point 2, which has a small distance from ice but smaller separation between first and second shell is found to have the same LSI value as point 3 which corresponds to a rather compressed and high density unicy environment. Although the effect is subtle, the LSI does better at separating undercoordinated waters from those that are overcoordinated which look very similar to non-defective waters.

\begin{figure}[!htb]
    \centering
    \includegraphics[width=15cm]{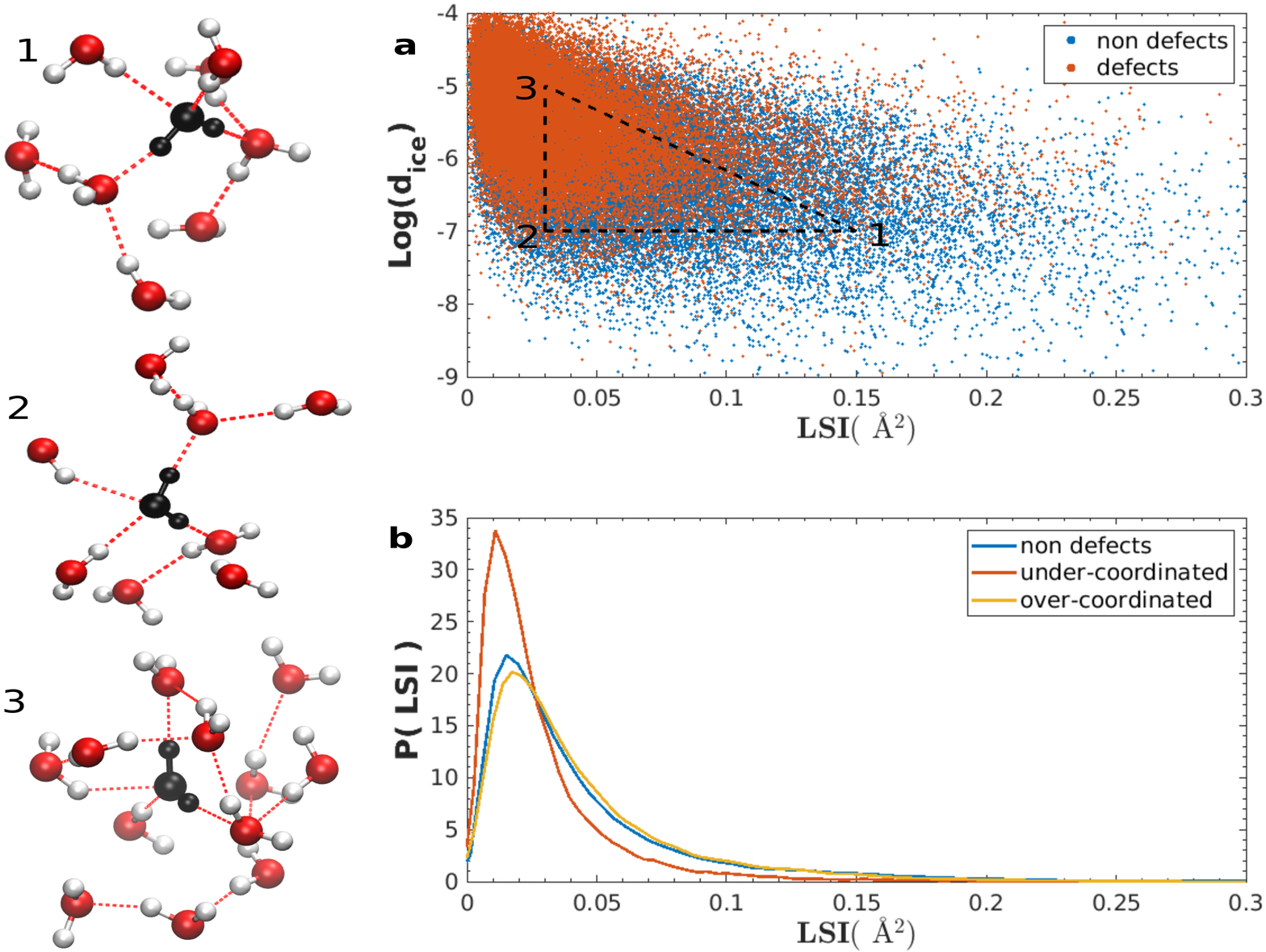}
    \caption{Panel a) shows the scatter plot of LSI versus Log($d_{ice}$) with the 3 numbered configurations corresponding to the environments shown pictorially on the left. Panels b and c show the probability densities obtained along $d_{ice}$ and LSI for the defective and non-defective water molecules.}
    \label{fig:lsi}
\end{figure}

% \begin{figure}[!htb]
%    \centering
%    \includegraphics[width=1.1\textwidth]{AliFigures/free3d.pdf}
%    \caption{Free energy surface constructed in 2D umap manifold colored by order parameters}
%    \label{fig:free3d}
%\end{figure} 

 %\begin{figure}[!htb]
 %   \centering
 %   \includegraphics[width=1.1\textwidth]{AliFigures/free3d_i%ce.pdf}
 %   \caption{free energy surface constructed in 2D umap manifold colored by distance from ice and free energy in SOAP space}
 %   \label{fig:free3d_ice}
%\end{figure} 

\subsection{Evolution of Molecular Descriptors with Free Energy}

In the preceding sections we have shown that the free energy landscape of liquid water at room temperature is best characterized as having one broad basin with small barriers separating the different minima. Furthermore, we have also seen as anticipated by the ID analysis, that the fluctuations within this landscape involve the coupling of several different molecular descriptors. In this section, we explore how these quantities change as a function of the free energy at room temperature. In addition, we also examine the behavior of some of the descriptors close to the critical point of supercooled water based on an analysis of microsecond long trajectories by Debenedetti and Sciortino\cite{debenedetti2020second}.

\subsubsection{Room Temperature Liquid Water}

Using the free energies of the points extracted earlier, we examined how the various descriptors, evolve as a function of being on different regions of the free energy surface. Figure 8 and Figure 9 show distributions of $q_{tet}$, d$_{5}$, LSI, $\rho_{vor}$ and Log($d_{ice})$ in slices of the free energy ranging between the minimum and 10 $k_{B}$T. Also shown in each panel, is the distribution of the respective variable obtained from all points independent of its position on the free energy surface (FES).

Starting with $q_{tet}$, we observe that the water
tetrahedrality reduces as one moves higher in FES. Interestingly, the shoulder at lower values of $q_{tet}\sim 0.5$ becomes much more pronounced for the points higher up in the FES and is consistent with the creation of more defects (see SI Figure \ref{fig:si_tetra_defects_slice}). The evolution of the variables such as $d_{5}$ and $\rho_{vor}$ reflect other changes in the hydrogen bond network. In particular, $d_{5}$ increases from 3.3 to $3.5$\AA{} moving above the minimum in free energy while $\rho_{vor}$ decreases from 0.037 to 0.03 \AA$^{-3}$. These changes correspond to water environments that become more open and less tetrahedral. It also worth noting that the points near the minima correspond to densities that are $12 \%$ larger than the average bulk density. The LSI distributions in Figure 8d shows more subtle changes toward higher values as a function of free energy again consistent with the formation of a more open local structure. It is clear however, that there is significant overlap along all these variables across the entire free energy landscape.

\begin{figure}[!htb]
    \centering
    \includegraphics[width=\textwidth]{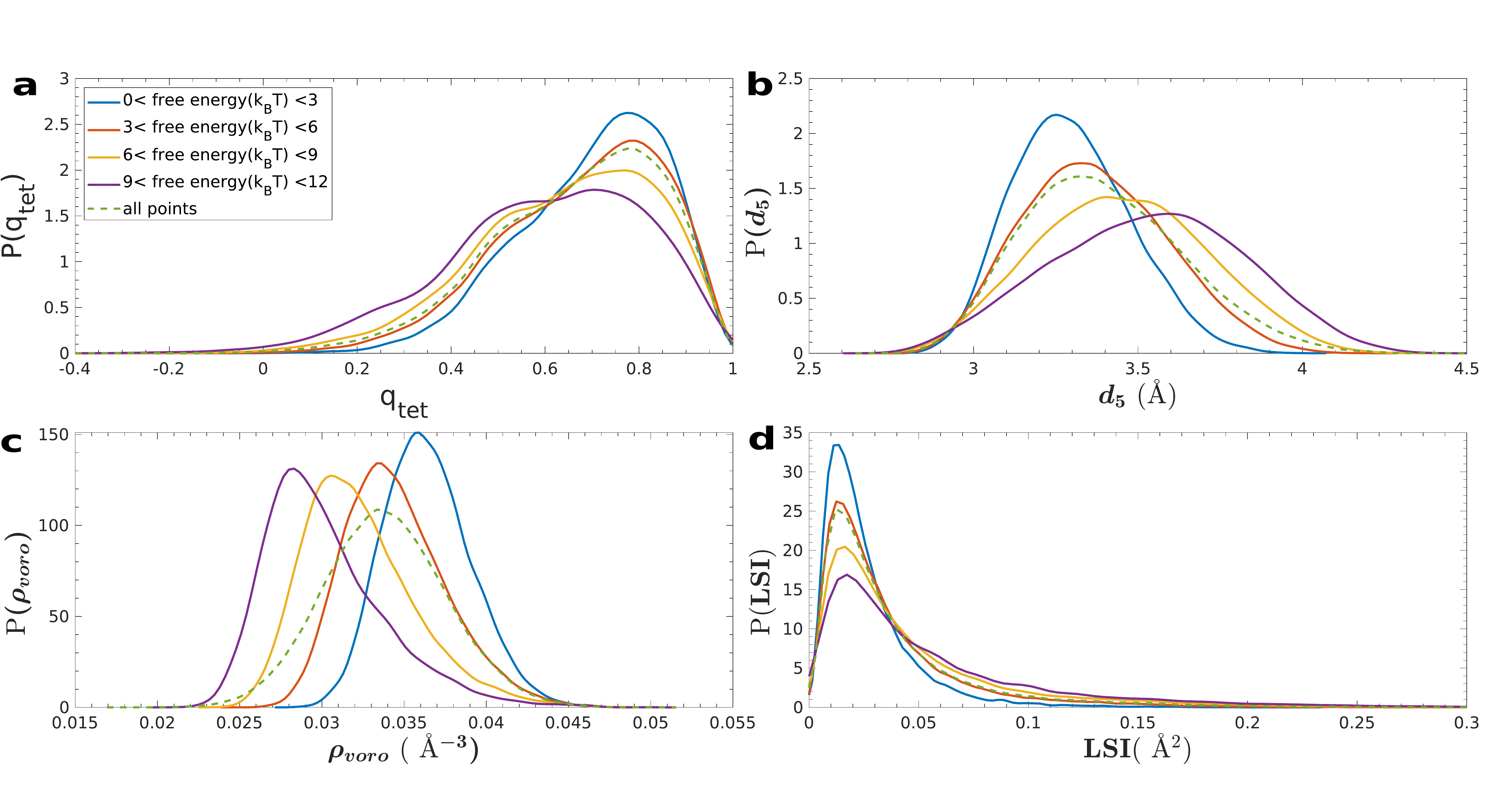}
    \caption{Panels a)-d) show the evolution of the chemical-based parameters as a function of different cuts along the free energy (shown with the various colored curves). The dashed curve in each panel corresponds to the distribution obtained by averaging over all the water molecules regardless of its free energy. }
    \label{fig:free_slice}
\end{figure} 

The left panel of Figure 9 shows the distributions of Log($d_{ice})$ for different free energy cuts. Interestingly, as one moves to regions of the FES that are higher in free energy, the environments look more ice-like which is consistent with the lower free energy structures in ambient temperature water being dominated by high density and more disordered environments. In the right panel of Figure 9, the changes in Log($d_{ice})$ as a function of free energy for both defective and non-defective water molecules are shown. Firstly, we note that the free energy minimum is characterized by the presence of both defective and non-defect water molecules consistent with the earlier analysis on the presence of a broad free energy basin characterized by low barriers separating different water structures. Secondly, we observe that fluctuations in the hydrogen bond network away from the free energy minimum results in the creation of both more ice-like or less-ice like environments as revealed by the changes in Log($d_{ice}$). For non-defects which accept and donate two hydrogen bonds, the higher lying free energy structures arise from more-ice like environments which are energetically stabilized but entropically disfavored.

Although the $\Log{(d_{ice})}$ we have used in this analysis only uses the the oxygen atoms, we have shown earlier that the inclusion of the hydrogen atoms contains important information as shown by ID (see Figure \ref{fig:scheme}) and SOAP-distance analysis (see SI Figure \ref{fig:si_inter_intra}). To see this a bit better within the context of $d_{ice}$, we compared the distribution of $\Log((d_{ice}$) for defects and non-defects comparing $\vec{\textbf{O}}$, ($\vec{\textbf{O}}$, $\vec{\textbf{H}}_{ave}$) and ($\vec{\textbf{O}}$, $\vec{\textbf{H}}_{ave}$,$\vec{\textbf{H}}_{dif}$). This analysis shows that when information of the environments of hydrogen atoms are included, $\Log(d_{ice})$ differentiates better between defects and non-defects.
(see Figure \ref{fig:si_def_non} and Figure \ref{fig:si_pdef}).

As eluded to earlier, recent theoretical studies proposed the possibility of fused dodecahedron structures as a possible source of a low density environment in water\cite{pettersson2019}. In order to assess this possibility, we also examined the $d_{dod}$ distance as described in the methods section. In the models of liquid water examined in this work, the environments in room temperature water yield larger values for $d_{dod}$ compared to $d_{ice}$ (see Figure \ref{fig:si_d_ice_slice}). 

%As we go higher up in free energy, we move from non-defective disordered environments towards either non-defective ordered environments or towards  non energetic defective environments

\begin{figure}[!htb]
    \centering
    \includegraphics[width=\textwidth]{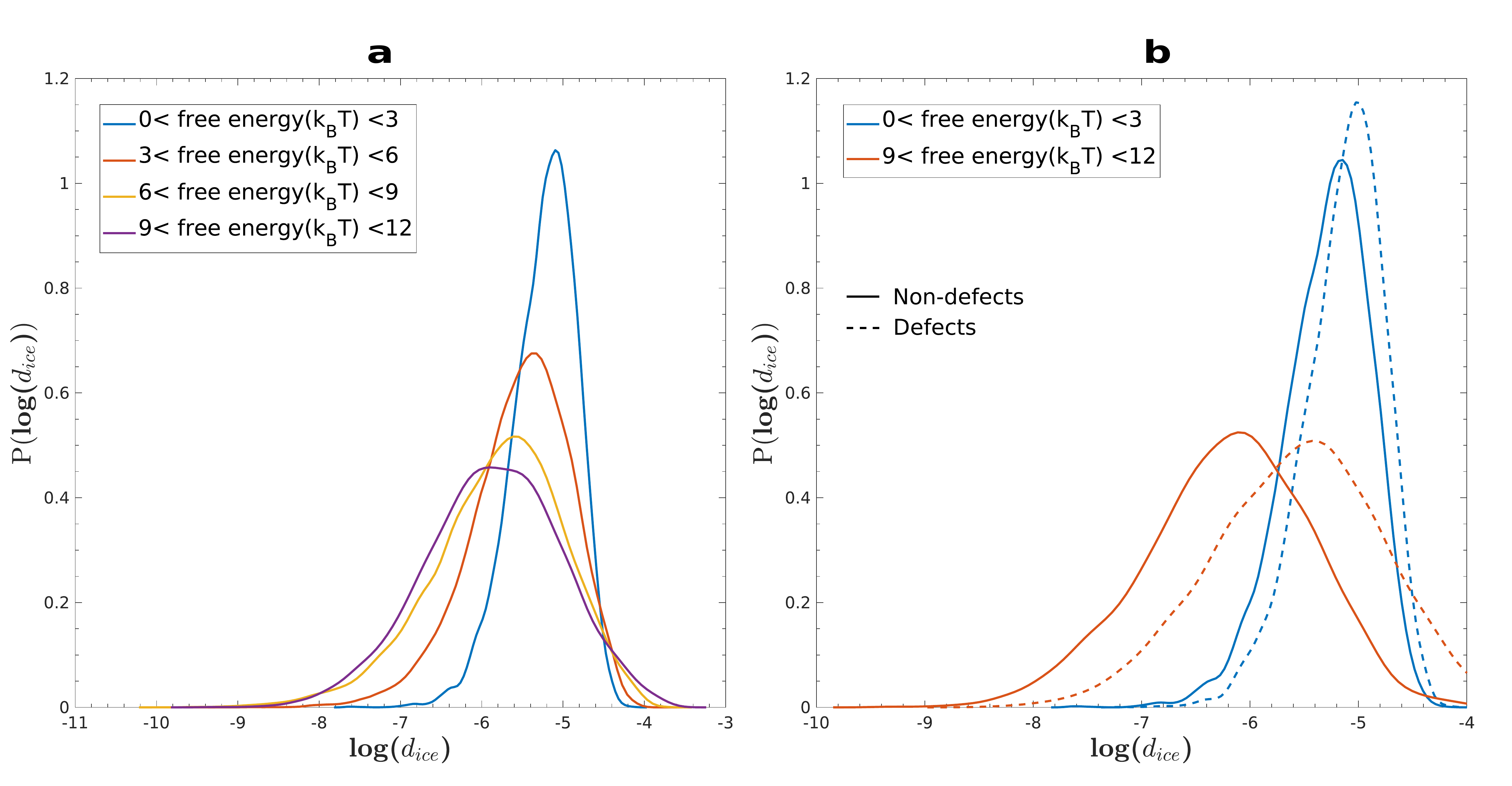}
    \caption{Panel a shows the evolution of $\Log(d_{ice}$) as a function of different free energy cuts while panel b illustrates the  $\Log(d_{ice}$)  for a low and high free energy region specifically for defective and non-defective water}
    \label{fig:d_ice_free}
\end{figure}

%\begin{figure}[!htb]
%    \centering
%    \includegraphics[width=\textwidth]{AliFigures/desc_free.pdf}
%    \caption{scatter plots of descriptors versus free energy at 300K }
%    \label{fig:scatter_desc_free}
%\end{figure}

\subsection{Supercooled Liquid Water}

\begin{figure}[!htb]
    %\centering
    \includegraphics[width=1.0\textwidth]{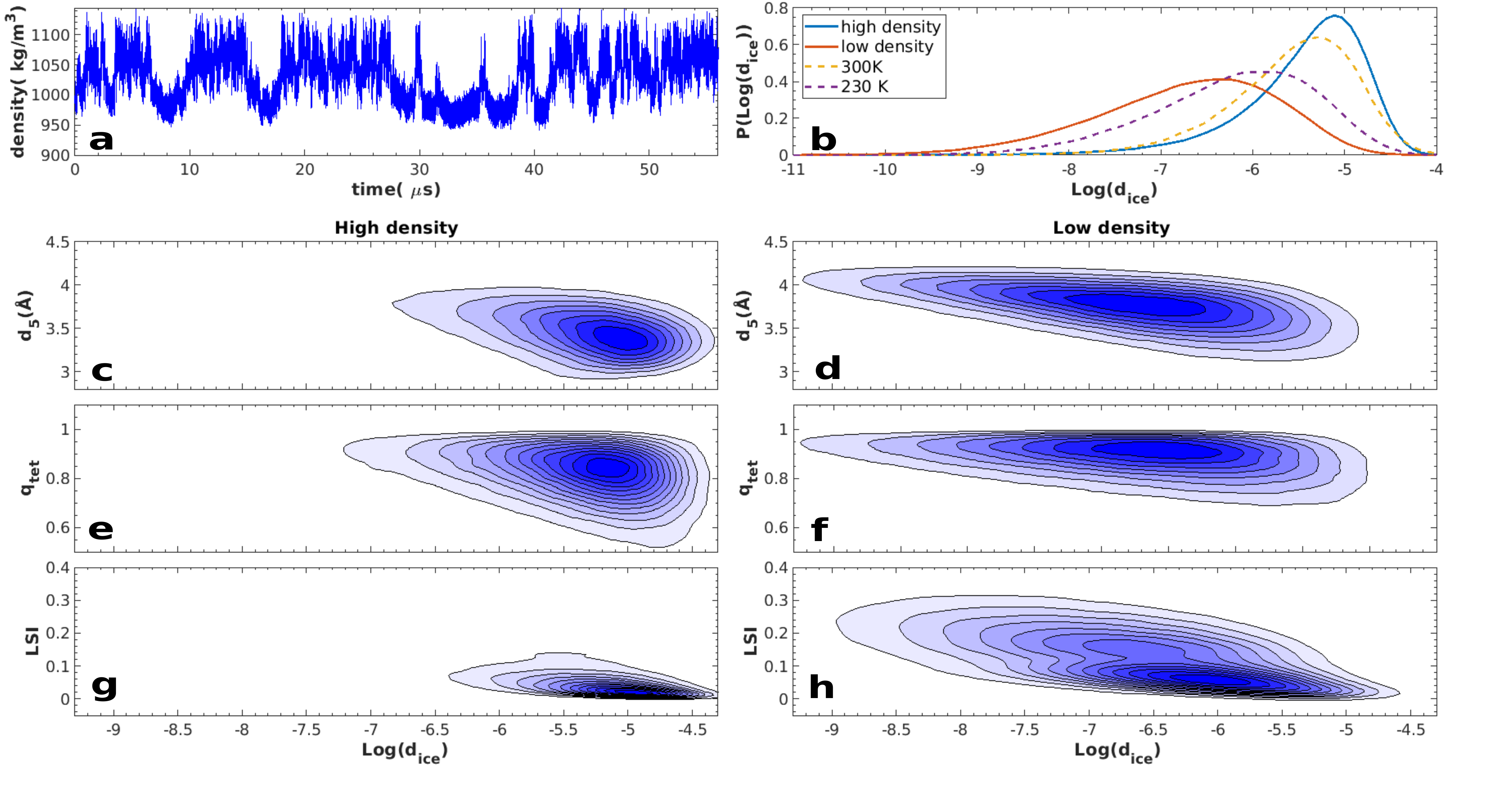}
    \caption{ Panel a is a plot of the time series of the global density of the trajectory at 171K,1751 bars. Panel b shows the comparison of the $\Log(d_{ice})$ at 3.7 \AA{}  between environments in  HD and LD regions contrasted with $\Log(d_{ice})$ at 230K, 300K. Panels c-h show the 2d kernel density estimates of the
$\Log{d_{ice}}$ versus chemical-based parameters.
Panels c and d, show the plots  d$_{5}$ and $\Log{d_{ice}}$ for HD and LD regions respectively while Panel's e and f  show those q$_{tet}$ and $\Log{d_{ice}}$. Figures g and h  show the plots LSI and $\Log{d_{ice}}$. The growth of a shoulder is observed in the LD phase for that the LSI}
    \label{fig:super}
\end{figure}

In a recent work, the second critical point of water was studied from long microsecond simulations of realistic point-charge water models\cite{debenedetti2020second}. In these simulations close to the critical point of water at $\sim 171K,1751 bars $, fluctuations between a high density (HD) phase at 1064 kg$m^{-3}$ and a low density (LD) phase at 977 kg$m^{-3}$ were observed. These transitions occurring over the course of several 10s of microseconds are illustrated in Figure \ref{fig:super}a . The local water environments of the HD and LD phases have typically been rationalized using chemical based descriptors such as $q_{tet}$, LSI and the bond-order Steinhardt order parameters\cite{appignanesi2009evidence,debenedetti2003supercooled,singh2016two,holten2014two,taschin2013evidence,gallo2016water,cuthbertson2011mixturelike}.% \ali{Adu: please fill in more references here}
%\alex{Adu, please, check the caption of Figure \ref{fig:super}}

We have seen earlier a combination of the chemical-based and SOAP variables provide a more nuanced perspective on the nature of the fluctuations in the hydrogen bonded network. In Figure \ref{fig:super} b, we show the $\Log(d_{ice})$ distributions for water environments extracted from the HD and LD regions of the trajectory in panel a. Specifically, SOAP environments were determined for all water in the frames where the density was within $1 kg/m^{3}$ of the minimum in the HD phase and LD phase separately. Additionally, the distributions for water at 300K and supercooled water at 230K are also shown. The LD phase is characterized by water environments that are more ice-like by 4-5 orders of magnitude compared to those in the HD phase. As expected, the environments observed in bulk water at 300K are much more similar to those in the HD phase close to the critical point. Interestingly, even though the global densities are quite different, there is a significant region of overlap in the environments observed in the HD and LD phases. This suggests the existence of heterogeneities within each of the liquid phases at the critical point.

Figure \ref{fig:super} c)-h) shows the behavior of the coupling between the $d_{5}$, $q_{tet}$ and LSI parameters as a function of $\Log(d_{ice})$ for the HD and LD phases in the left and right panels respectively. For both $d_{5}$ and $q_{tet}$ in panels c)-f), the extent of the correlation between these variables and Log($d_{ice})$ changes rather significantly when comparing the HD and LD phases. For the LD phase there appear to be a significant number of environments that have a large $d_{5}$ and high tetrahedrality, but cover a broad spread of Log$(d_{ice})$ values. The LSI for the LD phase is the only variable that shows the presence of a bimodal character. However in the HD phase, most of the local environments have the signatures of a closed, non-tetrahedral and higher local density (see SI Voronoi Figure \ref{fig:si_vor_hl}). A visual inspection of the trajectories suggests that the LD phase is characterized by the presence of larger domains that are built up of both low (where there are connected regions with lower $d_{ice})$) and smaller intermediately high density regions (with larger values of $d_{ice}$). Details of this analysis will be the subject of a forthcoming study\cite{adu2022superc}.

\section{Conclusions}

There is currently tremendous growth in the development and application of machine learning methods to understand complex molecular systems.
These approaches are aimed at circumventing the intervention of human or chemical bias as well as allowing for helping interpreting physical models that are beyond chemical imagination. 

In this work, we have used a series of advanced unsupervised learning techniques to study the fluctuations in simulated liquid water at room temperature. 
This procedure involves the use of state-of-the-art local atomic descriptors (SOAP) to describe water environments, followed by the extraction of the intrinsic dimension of the water network and then finally determining the topography of the free energy landscape. 
We also complement this analysis by studying the behavior of various chemically inspired coordinates that have been used to study water structure. 
We believe that this establishes a rigorous theoretical protocol for studying fluctuations in liquids and aqueous solutions in general.

Our analysis confirms previous several theoretical and experimental observations, that room temperature water is a homogeneous liquid.
However, the picture that emerges is much more nuanced. 
Fluctuations in the hydrogen bond network occur on a rather high dimensional free energy landscape that is broad and rather flat with small ripples separated by small free energy barriers. 
These features are found both in TIP4P/2005 water and in the many-body potential MB-pol water\cite{mbpol1}. This implies the presence of short-lived heterogeneities on the fs-to-ps timescale. The use of SOAP descriptors, by revealing the intrinsically multidimensional character of the local environment, leads to an additional conclusion: individually, variables such as $q_{tet}$, $d_{5}$, LSI or for that matter $d_{ice}$, cannot be used to infer the existence of LDL or HDL like environments. 

Finally, we also explore the evolution of all these variables within the supercooled regime by analyzing trajectories from recent work by Sciortino and Debenedetti\cite{debenedetti2020second}. 
While the HDL phase in supercooled water resembles the majority of local water environments in room temperature water, the situation is more complicated with LDL. 
Here there appear to be larger domains involving water environments that are more ice-like as well as high density-like but lower than the density of the HDL phase. 
The possibility of creating these domains is consistent with an earlier study by some of us showing that upon supercooling, a network of connected branched-voids develop surrounded by smaller spherical cavities\cite{ansari2019spontaneously}. 
Work is underway to apply the unsupervised learning approaches to characterize the free energy landscape of water near the critical point and will be subject of future work.

\begin{acknowledgement}
The authors acknowledge Francesco Paesani for sharing trajectories of MB-pol water, as well as Francesco Sciortino for sharing the supercooled water trajectories of TIP4P/2005. 
\end{acknowledgement}


\bibliography{achemso-demo}

\newpage

\end{document}

\begin{figure}[!htb]
    \centering
    \includegraphics[width=\textwidth]{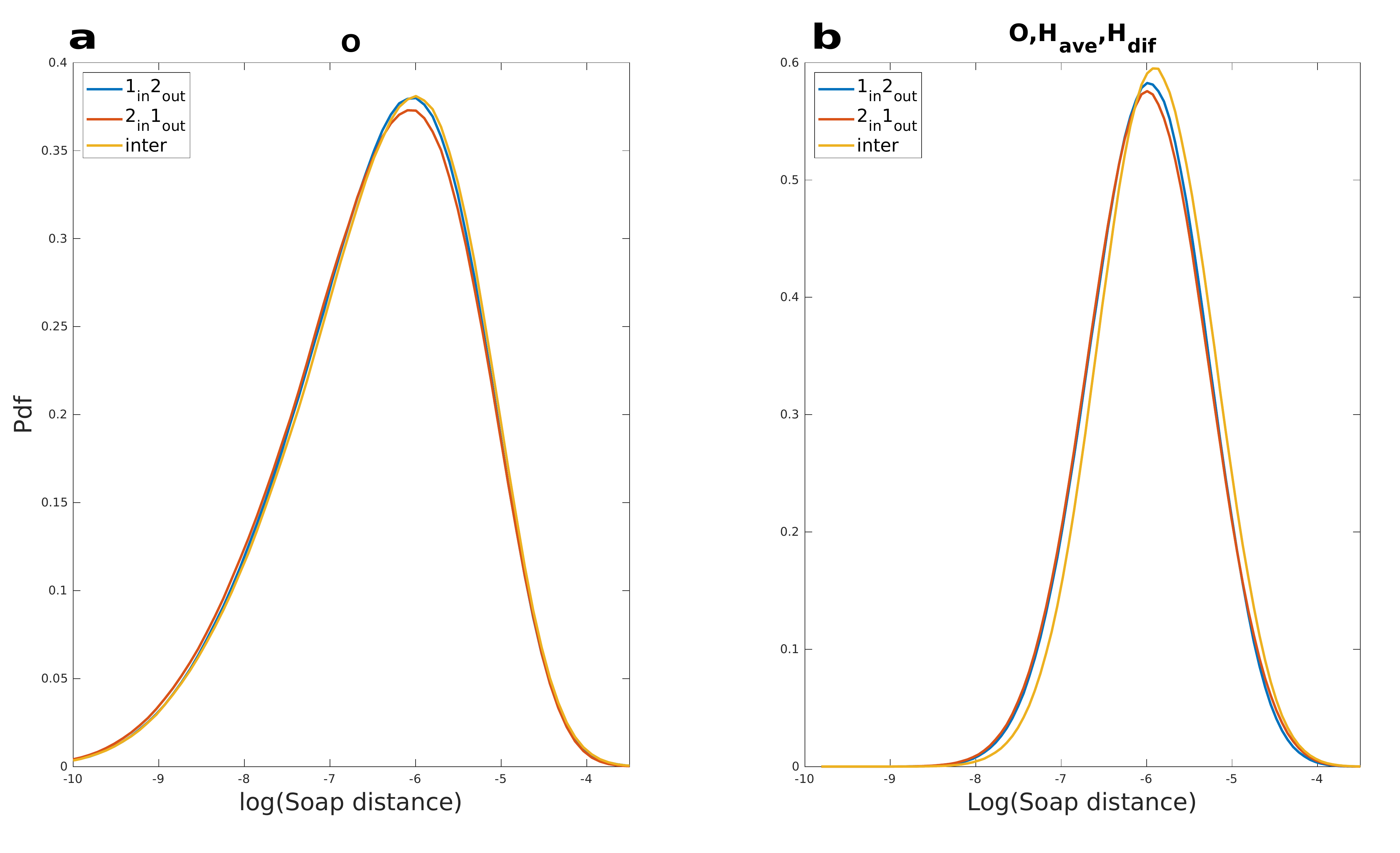}
    \caption{This figure shows the probability density estimates of the logarithm of SOAP distances between $1_{in} 2_{out}$ environments(blue) , $2_{in}1_{out}$ environments(red) as well as distances between $1_{in}2_{out}$ and $2_{in}1_{out}$ environments(yellow). Panel a shows that there is a complete overlap between all estimates when only oxygen atoms are used in computing the SOAP distances. In panel b however, we observe that the three distributions exhibit bigger differences when hydrogen atoms are included in the analysis.}
    \label{fig:si_inter_intra}
\end{figure}

\begin{figure}[!htb]
    \centering
    \includegraphics[width=\textwidth]{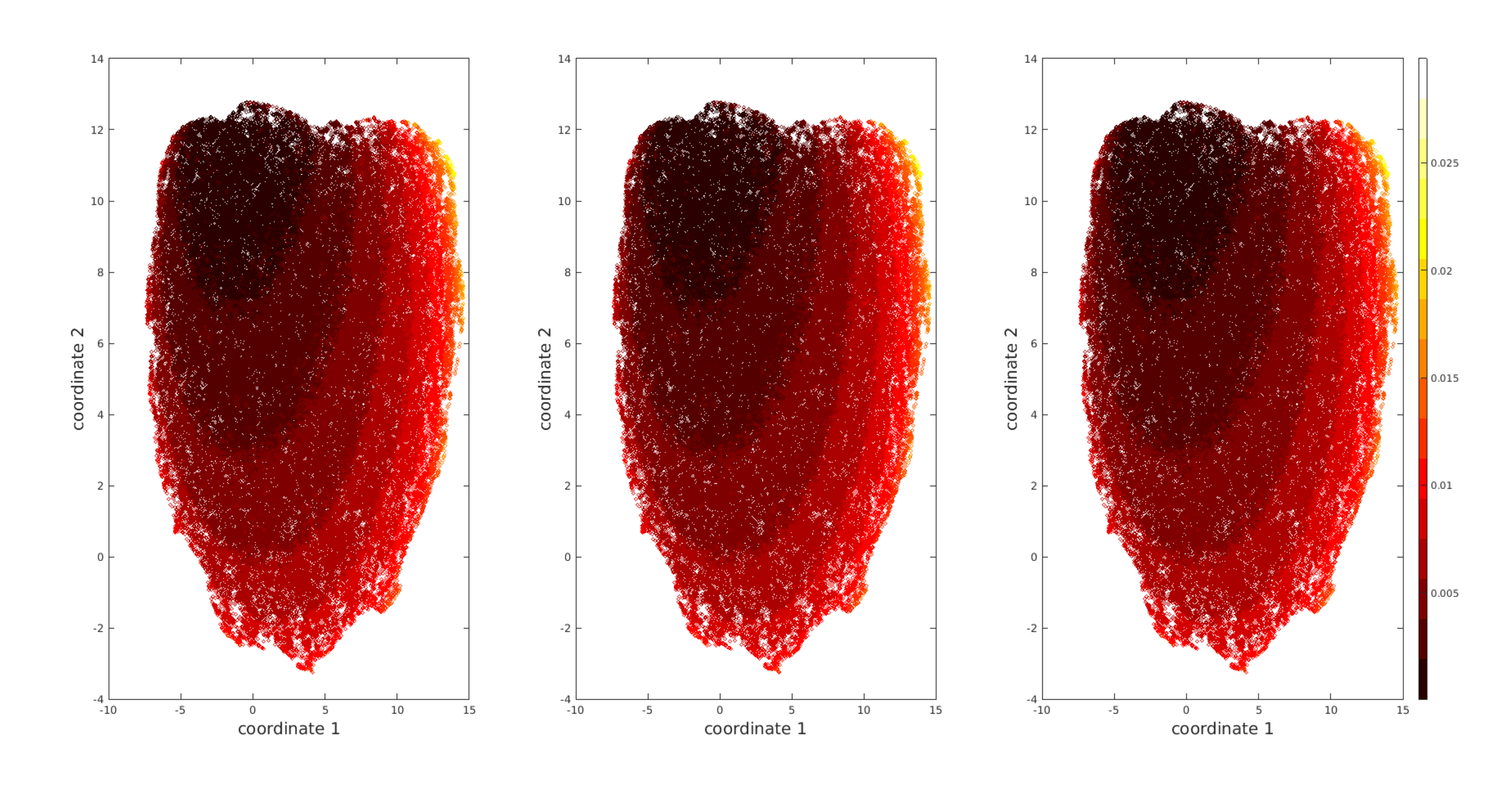}
    \caption{2D UMAP projection of the environments for three datasets colored by $d_{ice}$.}
    \label{fig:si_umap_3}
\end{figure}

%\begin{figure}[!htb]
%   \centering
%    \includegraphics[width=\textwidth]{AliFigures/hdbscan_tree_2.pdf}
%  \caption{This figure shows the condensed tree plot of the HDBSCAN for the three datasets}
%   \label{fig:si_hdbscan}
%\end{figure}

\begin{figure}[!htb]
    \centering
    \includegraphics[width=\textwidth]{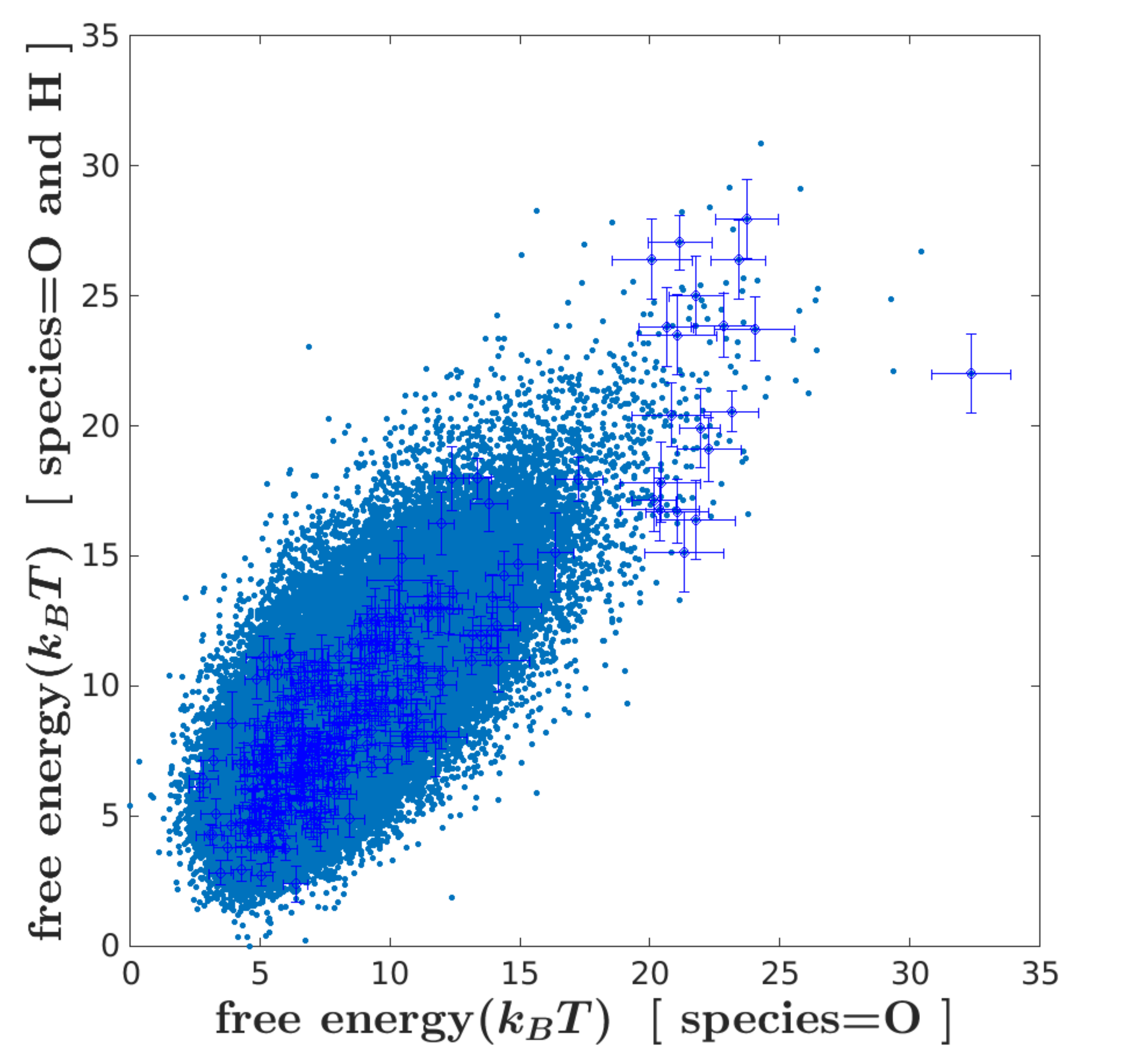}
    \caption{This figure which shows the scatter plot of the free energy values of $\vec{\textbf{O}}$ environments versus ($\vec{\textbf{O}}$, $\vec{\textbf{H}}_{ave}$,$\vec{\textbf{H}}_{dif}$). There is close to a linear relationship with a correlation coefficient of 0.7 and an RMSE between the two free energies of $\sim$2k$_{B}$T.}
    \label{fig:si_free_error}
\end{figure}

\begin{figure}
    \centering
    \includegraphics[width=\textwidth]{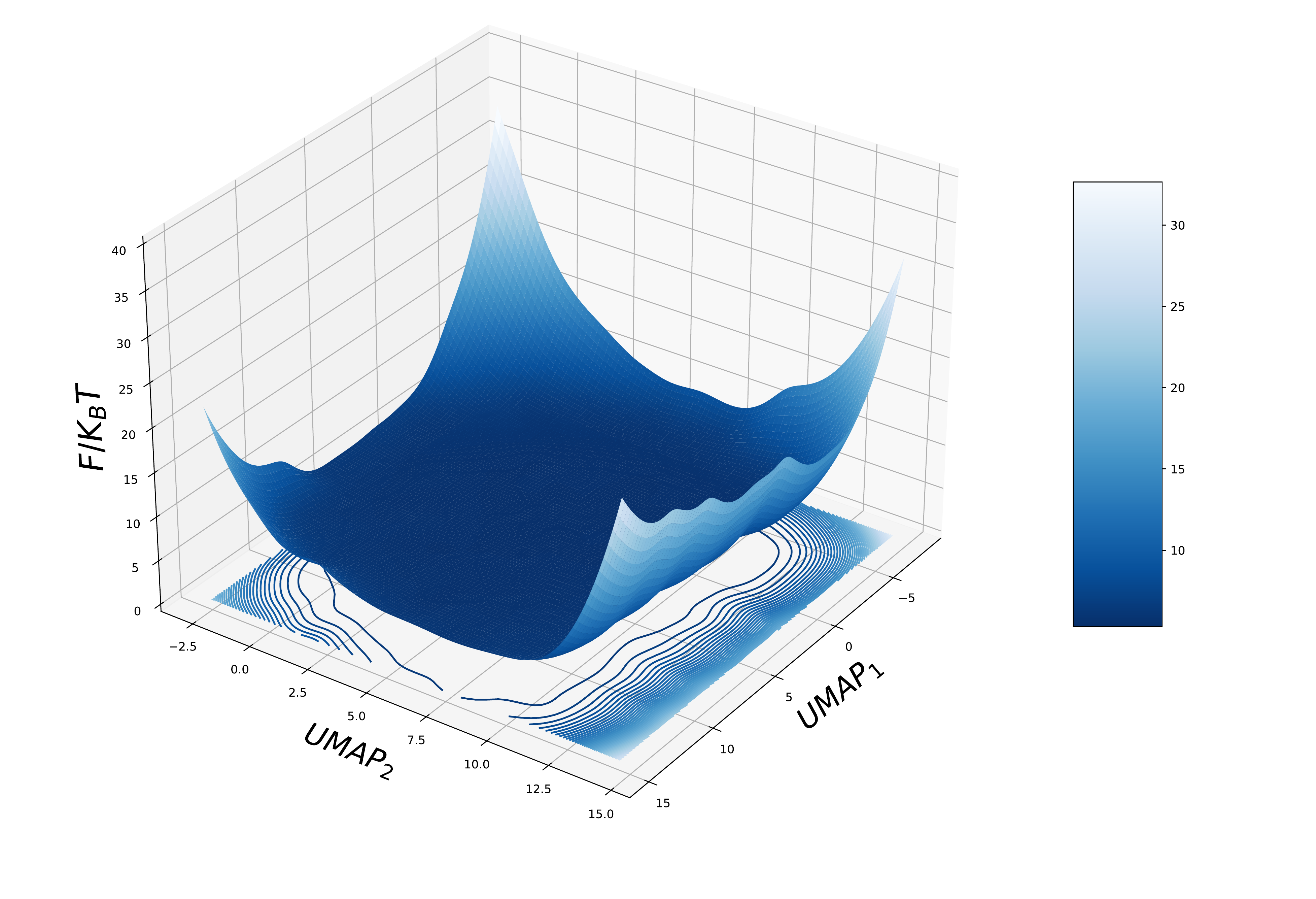}
    \caption{Free energy surface of MB-pol constructed in 2D UMAP manifold reveals a single basin without an appreciable barrier consistent with the results from TIP4P/2005 shown in the manuscript.}
    \label{fig:si_umap_mbpol}
\end{figure}

\begin{figure}[!htb]
    \centering
    \includegraphics[width=\textwidth]{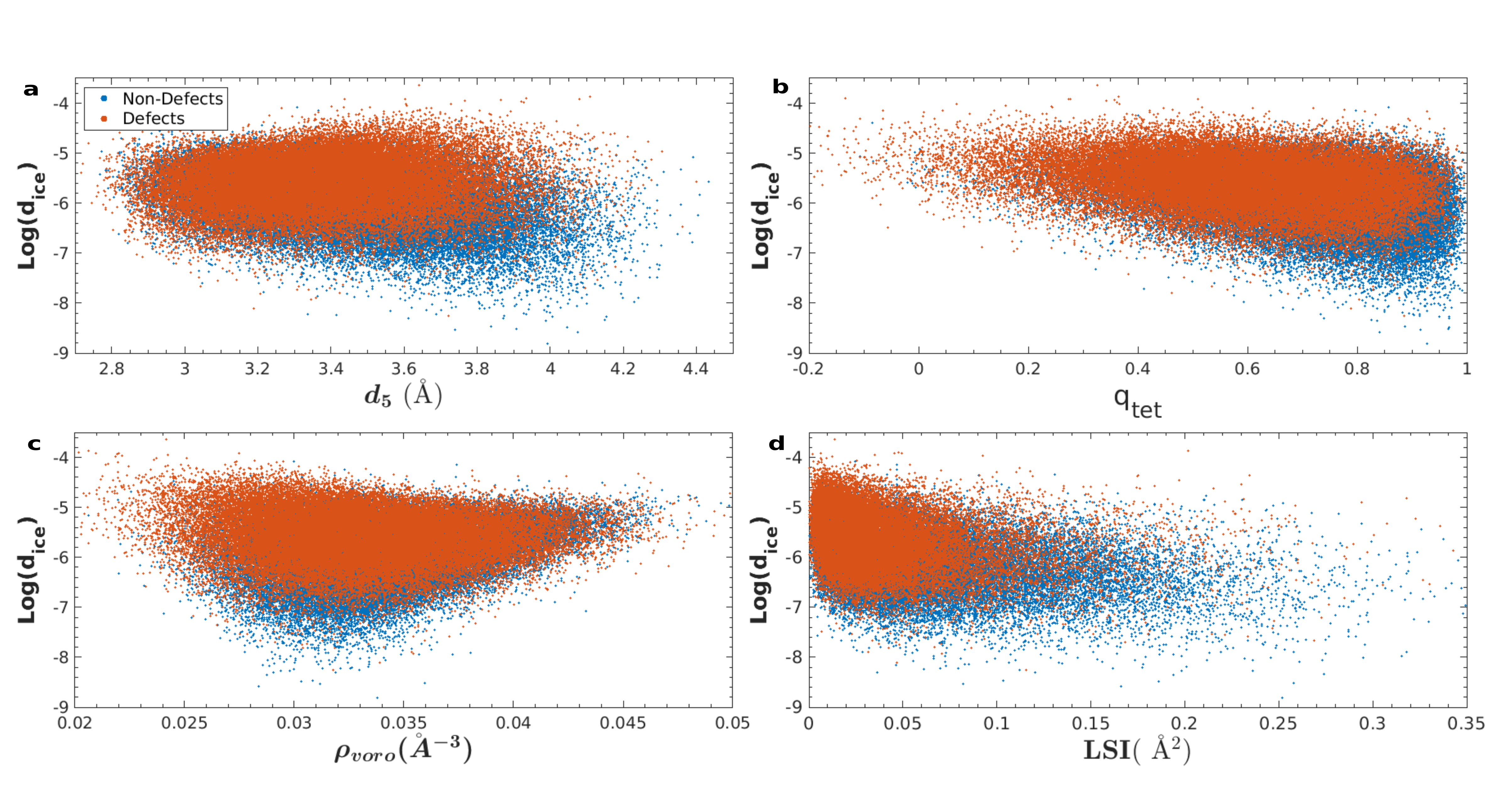}
    \caption{The panels a)-d) show the scatter plots of $\Log(d_{ice})$ at 3.7 \AA{} (including the hydrogen atom SOAP descriptors) versus the chemical-based collective variables for $q_{tet},d_{5},\rho_{voro}$ and LSI.}
    \label{fig:si_iceh}
\end{figure}

\begin{figure}[!htb]
    \centering
    \includegraphics[width=\textwidth]{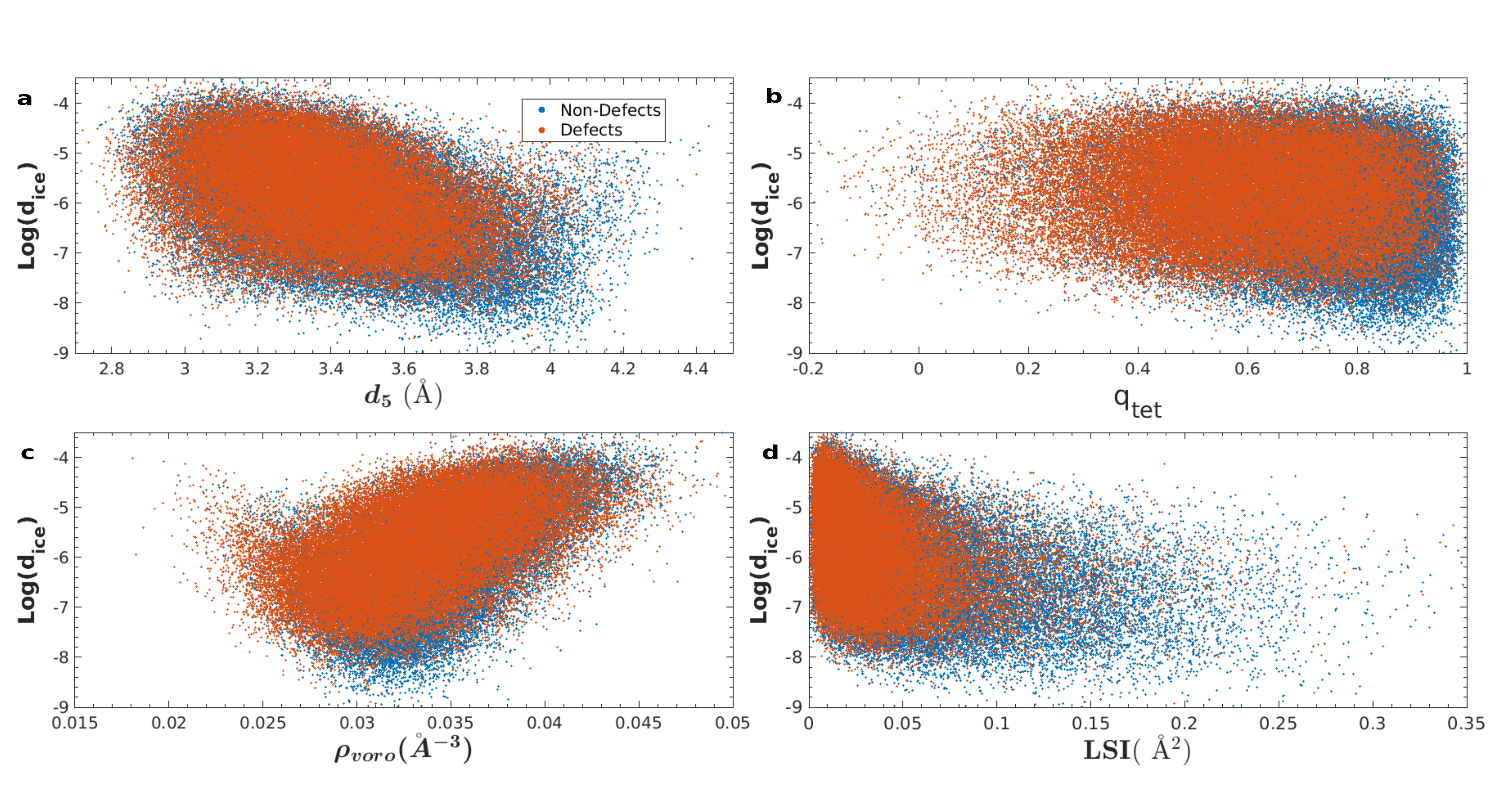}
    \caption{The panels a)-d) show the scatter plots of $\Log(d_{ice})$ at 6.0 \AA{} (including only the oxygen atoms for the SOAP descriptor) versus collective variables for $q_{tet},d_{5},\rho_{voro}$ and LSI.}
    \label{fig:si_ice6}
\end{figure}

   \begin{figure}[!htb]
    \centering
    \includegraphics[width=\textwidth]{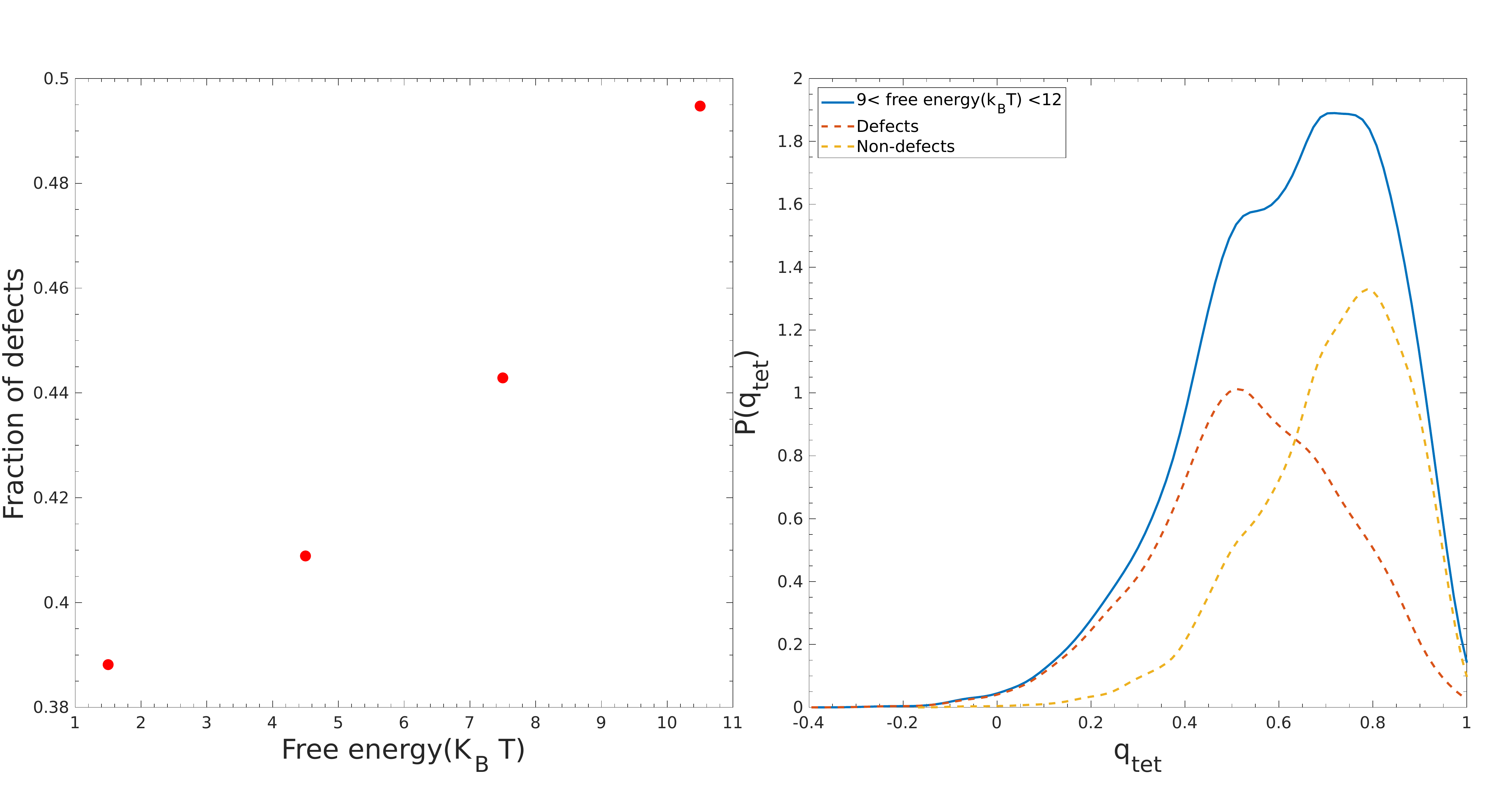}
    \caption{Left panel is the fraction of defects for 3$ k_{B}T$ cuts of the free energy. Right panel shows the $q_{tet}$ distribution of points high in free energy. Also shown are the weighted $q_{tet}$ distributions of points high in free energy restricted to defective(red) environments and non-defects(orange). }
    \label{fig:si_tetra_defects_slice}
\end{figure}

\begin{figure}[!htb]
    \centering
    \includegraphics[width=\textwidth]{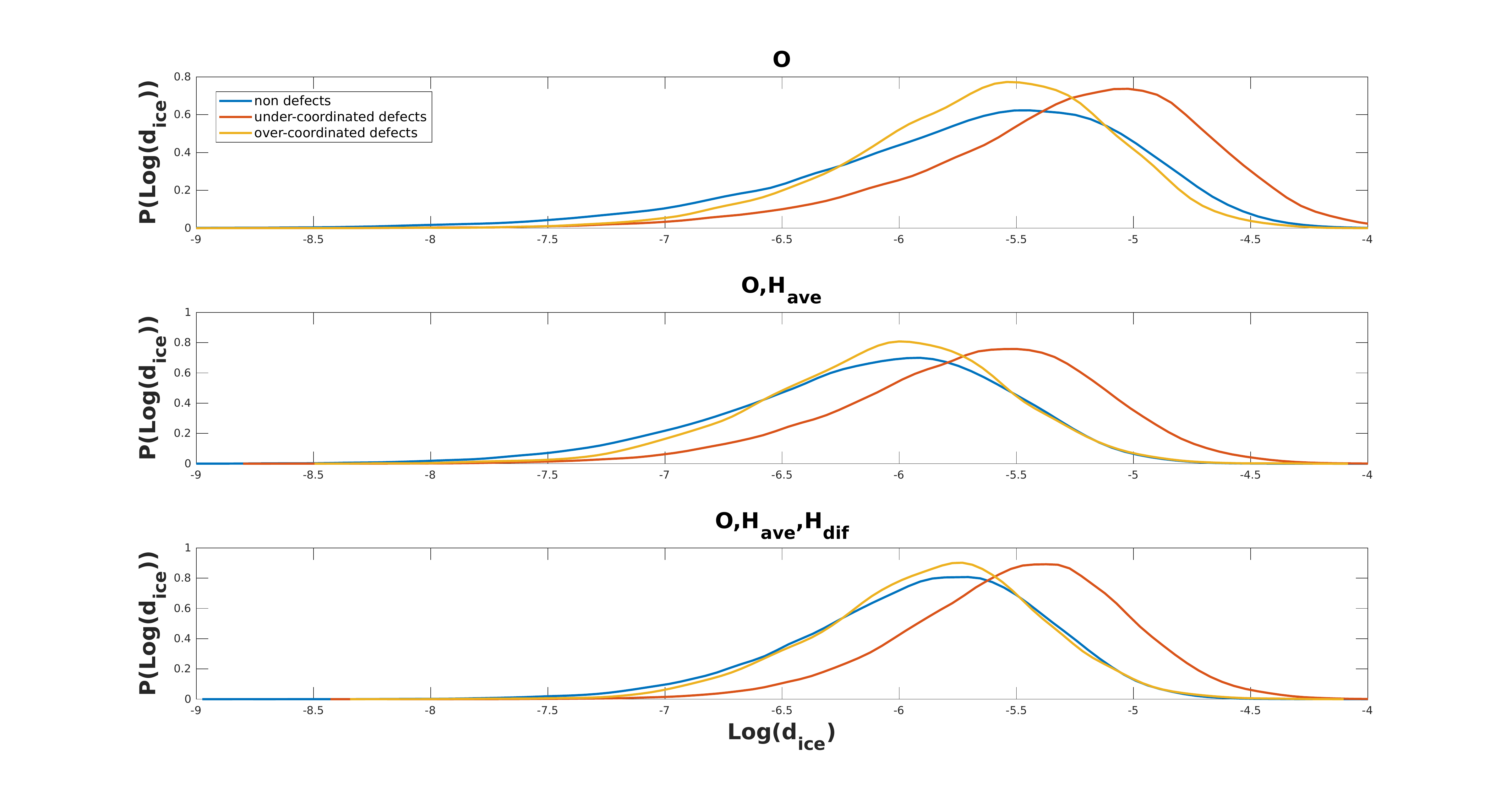}
    \caption{Probability density estimate of $\Log(d_{ice})$ for non-defects, under-coordinated defects and over-coordinated defects.}
    \label{fig:si_def_non}
\end{figure}

\begin{figure}[!htb]
    \centering
    \includegraphics[width=\textwidth]{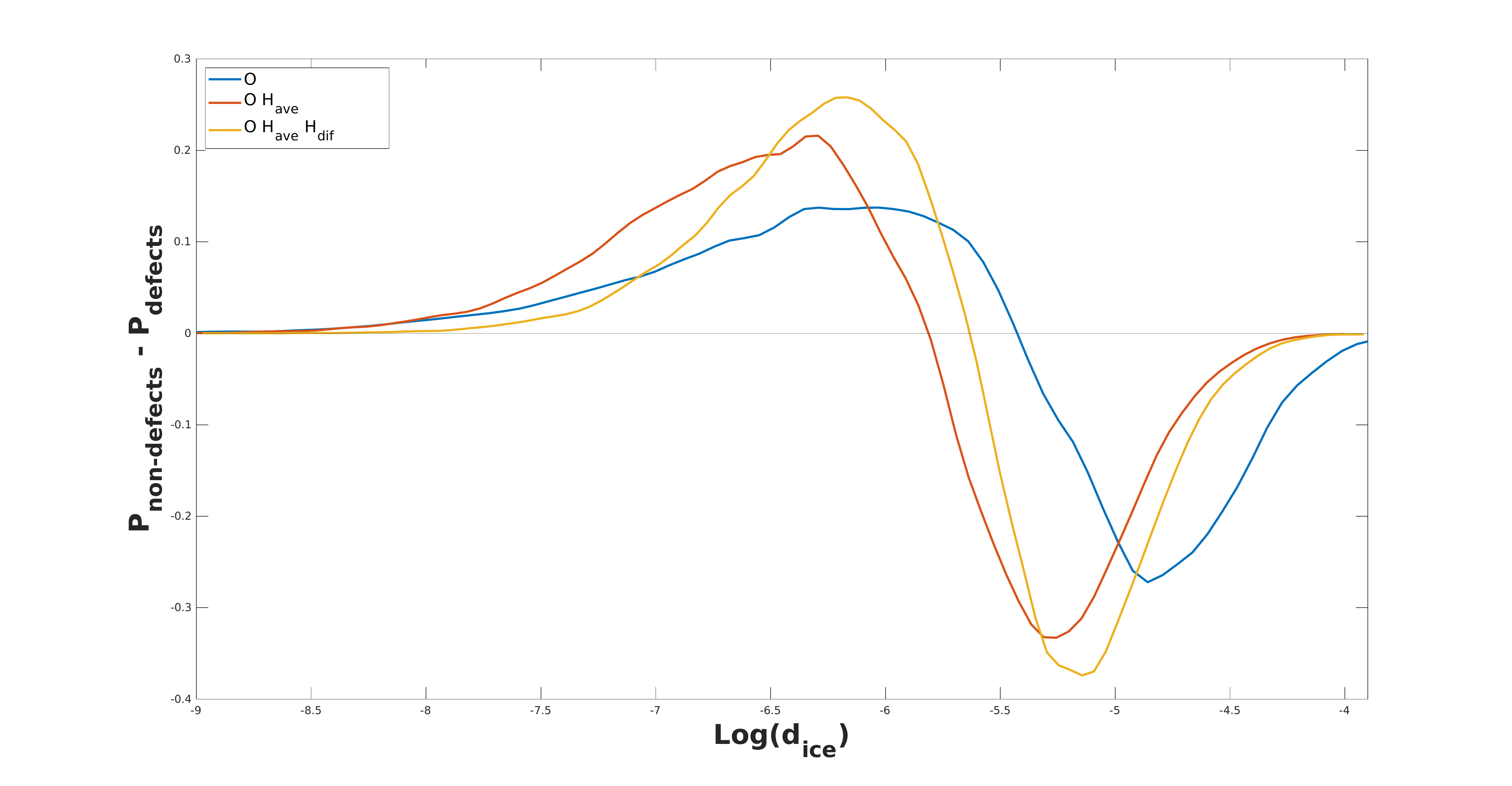}
    \caption{Figure shows the difference in distribution of $\Log(d_{ice})$ of non-defects and defects for the three variations of the SOAP descriptors: $(\textbf{O} )$, $(\textbf{O} \textbf{H}_{ave})$, $(\textbf{O},\textbf{H}_{ave},\textbf{H}_{dif})$. The descriptor including both $\textbf{H}_{ave}$ and $\textbf{H}_{dif}$ is found to have the greatest difference between defects and non-defects.}
    \label{fig:si_pdef}
\end{figure}

\begin{figure}[!htb]
    \centering
    \includegraphics[width=\textwidth]{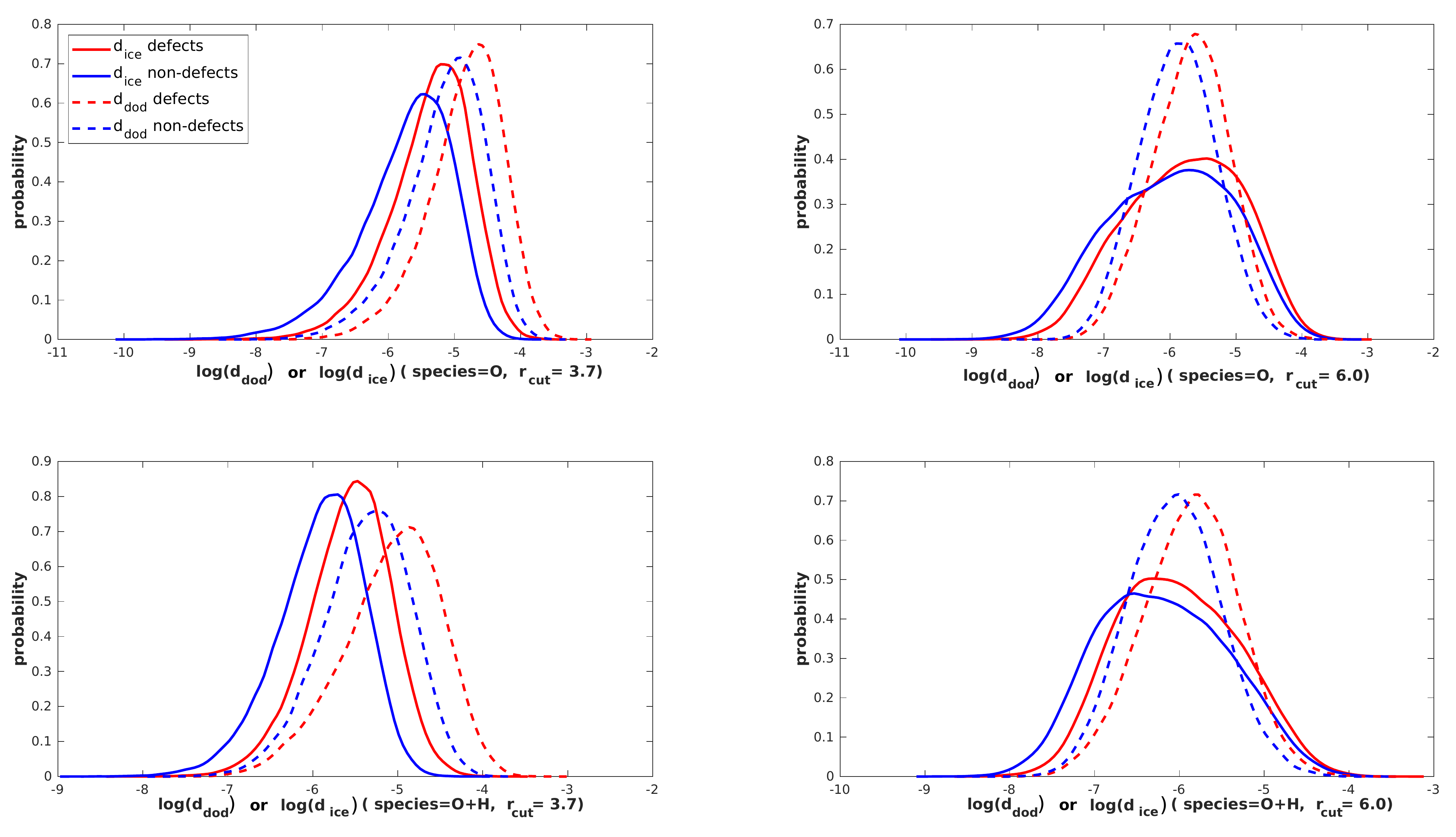}
    \caption{Probability density estimates of $\Log(d_{ice})$ and $\Log(d_{dod})$  restricted to defects and non-defects for radial cutoffs of 3.7 \AA{} and 6.0 \AA{}. The top panels are constructed using only oxygen atoms $(\textbf{O})$  while bottom panels include the hydrogen atoms in computing the distance $(\textbf{O},\textbf{H}_{ave},\textbf{H}_{dif})$.}
    \label{fig:si_d_ice_slice}
\end{figure}

\begin{figure}[!htb]
    \centering
    \includegraphics[width=\textwidth]{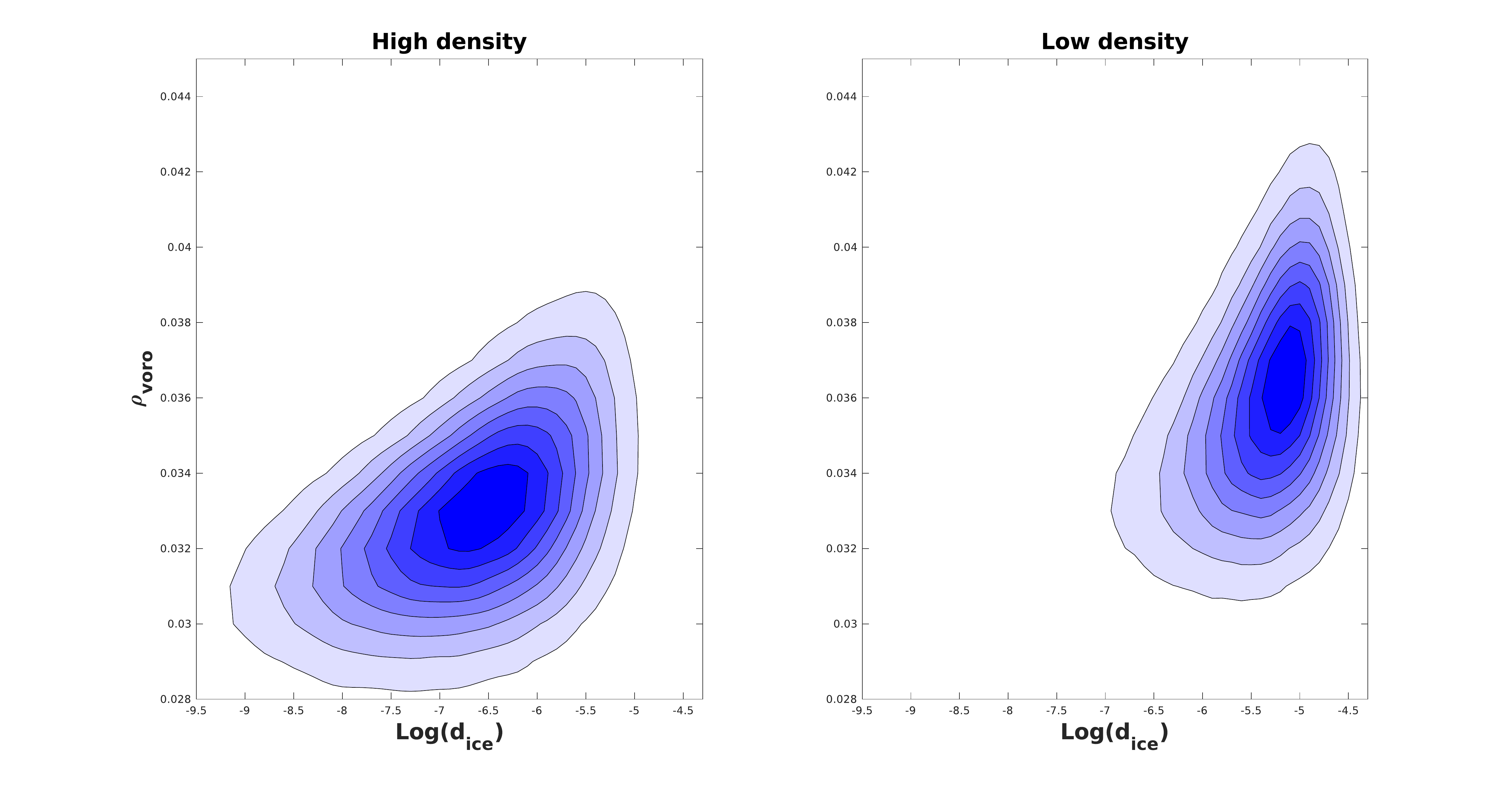}
    \caption{Density plot of $\log(d_{ice})$ versus $\rho_{voro}$ for HD and LD environments of supercooled water.}
    \label{fig:si_vor_hl}
\end{figure}